# Liquid structure adjacent to solid surfaces follows the superposition principle

Qian Ai[1,2], Haiyi Wu[3], Lalith Krishna Samanth Bonagiri[2,4], Kaustubh S. Panse[1,2], Shan Zhou[1,2], Fujia Zhao[1,2], Yitong Li[1,2], Kenneth S. Schweizer[1,2,5,6], Narayana R. Aluru[3] & Yingjie Zhang[1,2,7]*

1. Department of Materials Science and Engineering, University of Illinois, Urbana, Illinois 61801, United States

2. Materials Research Laboratory, University of Illinois, Urbana, Illinois 61801, United States

3. Walker Department of Mechanical Engineering and Oden Institute for Computational Engineering & Sciences, The University of Texas at Austin, Austin, Texas 78712, United States

4. Department of Mechanical Science and Engineering, University of Illinois, Urbana, Illinois 61801, United States

5. Department of Chemistry, University of Illinois, Urbana, Illinois 61801, United States

6. Department of Chemical and Biomolecular Engineering, University of Illinois, Urbana, Illinois 61801, United States

7. Beckman Institute for Advanced Science and Technology, University of Illinois, Urbana, Illinois 61801, United States

*Correspondence to: yjz@illinois.edu

**Summary paragraph:** Liquid structure at solid–liquid interfaces is critical for many natural and engineered processes ranging from biological signal transduction to electrochemical energy conversion[1,2]. Advanced experimental and computational methods have provided insights into the structure of liquids adjacent to planar substrates at the nanoscale[3,4]. However, realistic solid–liquid interfaces are inevitably inhomogeneous across multiple length scales, presenting a complexity that surpasses the capabilities of existing approaches[5]. Here we bridge the complexity gap by discovering and utilizing a hitherto hidden principle of interfacial liquid—superposition. Experimentally, we use 3D atomic force microscopy (3D-AFM) to image the interfacial structure of a wide range of organic and aqueous solvents and electrolytes, uncovering universal liquid density oscillations and emergent liquid layer reconfigurations at heterogeneous substrate sites. We further develop an analytical model, coined solid–liquid superposition (SLS), which solves the interfacial liquid density distribution based on a key descriptor: the effective total correlation function (ETCF) between a liquid molecule and nearby solid atoms. SLS not only explains all the experimentally observed interfacial liquid distribution profiles from the angstrom to near-micron scale, but also predicts more precise atomic-scale interference patterns which are further corroborated by molecular dynamics (MD) simulations. This study unveils a key structural descriptor of interfacial liquids, and establishes a theoretical framework for rapidly and accurately predicting liquid structures adjacent to solid surfaces with arbitrary morphology and size scale.

A major challenge in our understanding of solid–liquid interfaces is the complexity gap. Real-world interfaces in fields such as biology, renewable energy, and water purification are unavoidably heterogeneous across atomic, nano, micron, and larger scales[6,7]. Existing computational and experimental methods have enabled the accurate determination of the multiscale structure of solid surfaces[6,8]. However, our knowledge of interfacial liquids has been limited to those at atomically flat or nanoscale substrates[9,10]. Computational approaches face an accuracy–scalability trade-off: the most accurate simulations (e.g., ab initio molecular dynamics) are restricted to nanoscale systems[11], while scalable models (e.g., mean-field approaches such as the Gouy-Chapman-Stern theory[12]) are highly inaccurate within the crucial first 1–2 nm from the solid surface[11,13,14]. Experimentally, existing tools are constrained by the resolution–complexity trade-off: high-resolution imaging methods (e.g., 3D-AFM) face challenges when measuring complex interfaces with large roughness[2,9,10], while interface-sensitive spectroscopies (Raman, infrared, X-ray, etc.) can probe large, complex solid–liquid interfaces but lack spatial resolution[15–17].

To bridge the complexity gap, here we take a descriptor-based approach, with the goal of understanding and predicting the interfacial liquid structure through analytical formulas and parameters that are applicable regardless of the morphology and size scale of the adjacent solid. We first experimentally image the liquid structure at model solid surfaces with controlled heterogeneity. Based on the observed key features, we propose structural descriptors and integrate them into an analytical model, SLS, to explain the interfacial liquid structure. We then verify the predictive capabilities of SLS for arbitrary, complex solid–liquid interfaces.

**Experimental interfacial liquid structure at crystalline and atomic step sites**

We start by conducting 3D-AFM imaging (AC mode, amplitude modulation) of a model system, pure diethyl carbonate (DEC) solvent at highly oriented pyrolytic graphite (HOPG) surface. This system both mimics the negative electrode–electrolyte interfaces in lithium-ion batteries and hosts a series of well-defined substrate heterogeneities[18]. During 3D imaging, the AFM cantilever's amplitude and phase are recorded as a function of x, y, and z, from which we construct the background-removed conservative force ($\Delta F$)[2,19–21] (see Methods). At flat, crystalline basal plane areas, we observe damped oscillations in the $\Delta F$ vs z (vertical distance from the HOPG surface) profile (Fig. 1a,b). The overall damped oscillation pattern is consistent with previously reported interfacial liquid structure of pure water, aqueous solutions, organic electrolytes, and ionic liquids on flat substrates, as validated by 3D-AFM, X-ray scattering, and/or atomistic simulations[19]. Liquid density oscillation near solid surfaces is a universal phenomenon, arising from the finite size of the liquid molecules together with the entropic constraints imposed by the substrate[2,19,22,23]. The oscillation period of DEC/HOPG is found to be nearly constant at ~4.0 Å (Fig. 1b), which corresponds to the center-to-center distance of adjacent DEC molecules[24].

To investigate the impact of surface heterogeneity on the nearby liquid structure, we scan the HOPG surface and identify a series of clean atomic step sites, including mono-step and multi-step edges (Extended Data Fig. 1). 3D imaging is then performed at these sites. In the 3D-AFM community, both the local phase minima and force maxima have been conventionally used as indicators of liquid layer positions, with phase having higher signal-to-noise ratio and force being more quantitative[2,9,19–21]. From the x-z force/phase maps (Fig. 1c–e, Extended Data Fig. 2, and Supplementary Figs. 1–3), we observe remarkable layer transitions across the step edge areas, with two key features: 1) The liquid layers above the upper HOPG surface (left of the step edge in Fig. 1c–e) "slip" at the step edge, connecting to a nearby liquid layer above the lower HOPG surface; we name this effect as "layer crossover", as the liquid layer number changes across the step edge. 2) Above the lower surface, the liquid layers at z positions lower than that of the upper HOPG surface terminate near the step edge, although the exact termination spot cannot be determined in the bi-step and tri-step systems due to the finite size of the AFM tip (Extended Data Fig. 3 and Supplementary Note 1). Overall, the liquid layers exhibit non-conformal coverage near the step edge site—they do not follow the topographic contour of the substrate.

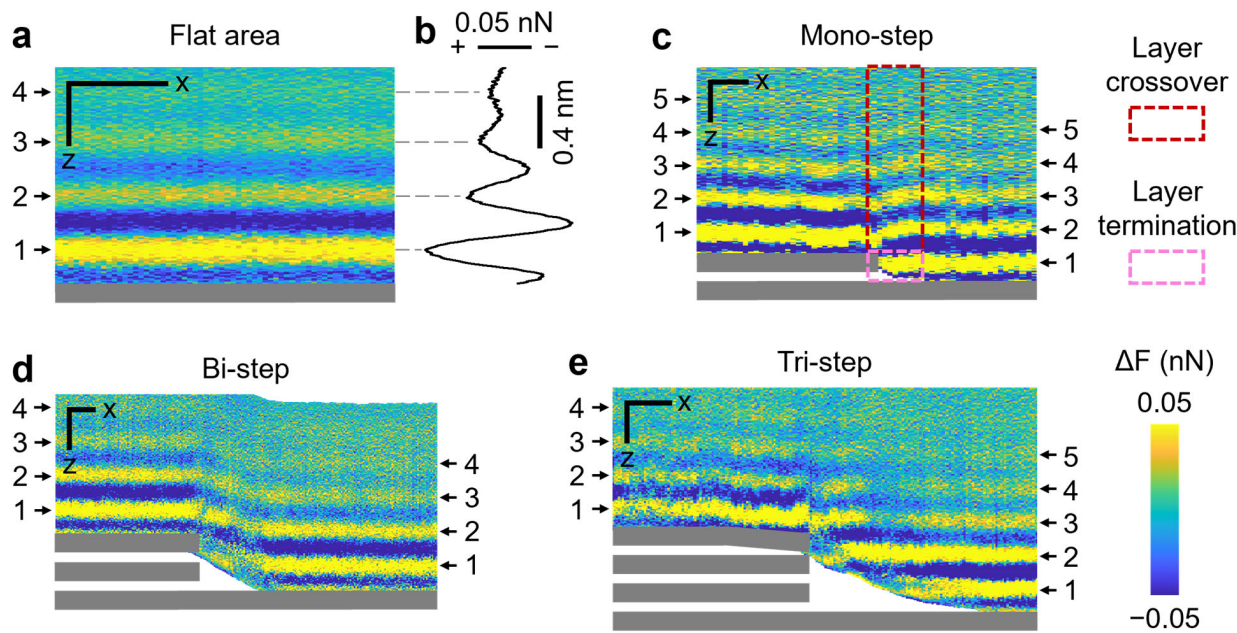

**Fig. 1 | 3D-AFM force maps at different sites of the DEC/HOPG interface. a**, Flat area x-z force (ΔF) map. HOPG surface is marked by the gray line at the bottom, while DEC liquid layers are labeled by the corresponding layer numbers. **b**, Force vs z profile by averaging along the x direction in **a**. **c–e**, 3D-AFM x-z force maps of DEC near HOPG atomic step sites. Each gray line represents a graphene layer as part of the HOPG substrate. Step heights are observed as 0.34 nm, 0.68 nm, and 0.94 nm for mono-step (**c**), bi-step (**d**), and tri-step (**e**), respectively. DEC layers on each side of the step edge are labeled by the corresponding layer numbers, highlighting the layer crossover effects. In **c**, the regions exhibiting layer crossover and termination effects are also marked by the dark red and pink dashed rectangles, respectively, to guide the eye. In **d** and **e**, the blank regions (white, triangular-shaped) adjacent to the step edges correspond to areas the AFM tip cannot access due to its finite radius (Extended Data Fig. 3 and Supplementary Note 1). In **e**, the top layer is bent downward near the step edge, which is common for graphitic structures (see more discussions in Supplementary Note 2). Scale bars: 3 nm for x and 0.5 nm for z (**a,c–e**).

In prior reports from our lab, both conformal and non-conformal interfacial liquid layers have been observed for aqueous solutions and lithium salt/ionic liquid electrolytes on substrates with heterogeneous adsorbates or clusters[2,25]. Other labs have also reported 3D-AFM imaging of aqueous solution at non-flat substrates[26–28]. However, due to the randomness of the surface corrugation and/or the limited resolution/sensitivity, the underlying mechanism has been elusive. Here the DEC layers are resolved at the well-defined atomic step sites of HOPG. The layer crossover effect is identified unambiguously. For example, layer 1 on the left crosses into layer 2 on the right for the mono- and bi-step systems; for the tri-step system, layer 1 crosses into layer 3 from left to right (Fig. 1c–e). Within the connected liquid layers, we also observe a shift of force values across the step edge, further confirming the layer crossover effect (Supplementary Fig. 4). The finite AFM tip radius has impacts on the experimentally observed transition region width of both the substrate and the liquid layers, but does not affect the qualitative layer crossover features (Extended Data Fig. 3 and Supplementary Note 1).

We have measured DEC/HOPG (flat area and step edge) systems through five separate sets of experiments using two different types of cantilevers. Each experiment was conducted with a new AFM tip having a specific (likely different) tip radius. In total, we have obtained more than 16,000 x-z maps and >1 million force curves. The force curves are highly reproducible at all the flat area sites (Extended Data Fig. 4), while the layer crossover and termination effects are consistently observed at atomic step sites in all the measurements (Fig. 1c–e, Extended Data Fig. 2, and Supplementary Figs. 1–3).

The observed non-conformal effect indicates that the interfacial liquid structure cannot be predicted by simply following the surface corrugations of the substrate. To unravel the universal relationship between substrate morphology and interfacial liquid structure, a new analytical framework is needed—one that extends beyond models developed for flat solid–liquid interfaces.

**Solid–liquid superposition model**

To capture the key mechanism of the interfacial liquid structure, we take a descriptor-based analytical approach. We consider the exact atomic structure of the solid, and parameterize the solid-modulation of adjacent liquid density via an ETCF, $h_{eff}(\boldsymbol{r}_{sl})$. Here, $\boldsymbol{r}_{sl} = \boldsymbol{r}_l - \boldsymbol{r}_s$, with $\boldsymbol{r}_l$ representing the geometric center of a liquid molecule and $\boldsymbol{r}_s$ the location of a solid site (an atom, lattice, or other repeating unit). At this stage, we focus on the liquid density profile, as represented by the spatial distribution of the molecules' geometric centers. This simplification is universally applicable to liquid molecules regardless of their shape, with the limitation that the molecular orientational information is not captured.

The real total correlation function (TCF) between the liquid molecules and a solid site represents the local real-space liquid density profile. At heterogeneous sites where translational symmetry is broken, TCF is subject to change at different sites. Therefore, TCF cannot serve as a universal descriptor of the interfacial liquid structure. To overcome this limit, here we define ETCF to capture the effective contribution of individual solid sites to the variation in nearby liquid density. For the same solid site–liquid molecule pair, ETCF is identical, regardless of the local solid morphology. Therefore, ETCF can serve as a powerful descriptor for deriving the interfacial liquid structure.

For example, in the DEC/HOPG system, if we take a carbon atom (of HOPG) as a site, the ETCF between the carbon site and nearby DEC molecules is identical everywhere at the interface. This includes the flat basal plane areas, step edges, and other heterogeneous sites where only carbon atoms are present. If impurity atoms (e.g., oxygen and nitrogen) exist, different ETCFs may be needed. In this work, we mainly consider systems where only one ETCF is required to describe the key interfacial liquid profiles. As shown in the following results, such systems can be highly heterogeneous and even include multiple electrolyte components. For systems requiring more than one ETCF, our developed analytical model can also be readily generalized to obtain their liquid density distributions.

A bulk liquid in equilibrium has a uniform time-averaged density, $\rho_{bulk}$, due to the mobile nature of the liquid species. For a given ETCF, we propose that the time-averaged, near-surface liquid density variations can be obtained through a linear superposition:

$$\frac{\Delta\rho(\boldsymbol{r}_l)}{\rho_{bulk}} = \sum_{r_s} h_{eff}(\boldsymbol{r}_{sl}), \tag{1}$$

where $\Delta\rho(\boldsymbol{r}_l) = \rho(\boldsymbol{r}_l) - \rho_{bulk}$, and $\rho(\boldsymbol{r}_l)$ is the liquid density at position $\boldsymbol{r}_l$.

Eq. 1 is the core of our proposed SLS model. The profound impact is, as long as the solid atom/site distribution and ETCF are known, in either analytical or numerical form, we can instantaneously calculate the liquid density distribution adjacent to a solid with arbitrary surface morphology. $h_{eff}(\boldsymbol{r}_{sl})$ can be obtained via multiple alternative approaches, such as analytical models, MD simulations, and/or experimental inputs (e.g., 3D-AFM and X-ray scattering).

In the following, we show how ETCF can be quantified by integrating an analytical formula with either 3D-AFM data or MD simulation results. Based on classical theory of liquids, the TCF of pure liquids and liquid mixtures are usually in the form of damped oscillations, due to the dense packing of the discrete molecules[29–31] (Supplementary Note 3). Considering the generic oscillations and the impact of substrate–liquid interactions, we propose an analytical form of ETCF:

$$h_{eff}(\boldsymbol{r}_{sl}) \approx A_0 \exp(-\alpha_2 r_{sl}) \cos(\alpha_1 r_{sl} - \theta_0)/r_{sl} + A_1 \exp\left[-\frac{(r_{sl} - r_1)^2}{2\sigma^2}\right], \tag{2}$$

where the first term is hypothesized based on predictions of the Ornstein-Zernike (OZ) equation[32], and the second term is an empirical correction of the first liquid density peak. $A_0$, $1/\alpha_2$, $2\pi/\alpha_1$, and $\theta_0$ correspond to the amplitude, decay length, periodicity, and phase factor for the damped oscillation function, respectively. $A_1$, $r_1$, and $\sigma$ represent the amplitude, center and width of the Gaussian function for first peak correction, respectively (see Supplementary Note 3 for more details).

Combining experimental 3D-AFM force maps of the DEC/flat HOPG interface with Eqs. 1 and 2, we extract parameters for the ETCF between the DEC molecule and carbon atom of HOPG (Supplementary Notes 4–6 and Supplementary Table 1). The obtained ETCF, shown in Fig. 2a, exhibits strong damped oscillations with a periodicity of 4.0 Å, consistent with the 3D-AFM results. We then construct the atomic configurations of HOPG substrates, including both the basal plane and step edge areas (Fig. 2b, bottom and Supplementary Figs. 5–7). Substituting the carbon atomic positions ($r_s$) and the DEC–carbon ETCF ($h_{eff}(r_{sl})$) into Eq. 1, we calculate the 3D DEC density maps at the interface (Fig. 2b–i). We refer to this analytical calculation method as "AFM-SLS", since it uses 3D-AFM data to extract the ETCF descriptor and then employs the obtained ETCF for SLS calculations (see summary in Supplementary Fig. 8).

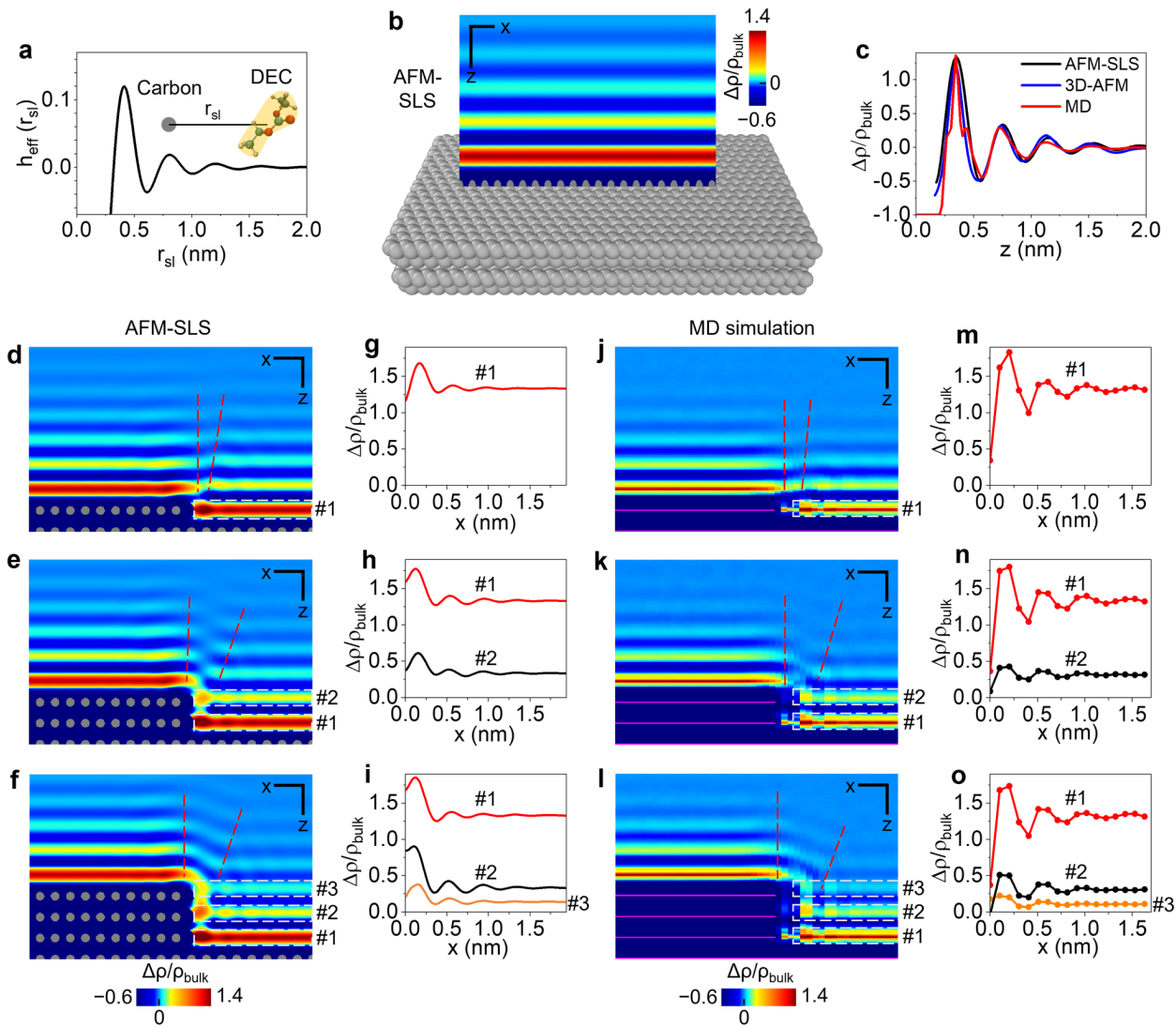

**Fig. 2 | SLS model of interfacial liquid structure: key features and benchmarking.** System: DEC/HOPG interface. **a**, The ETCF for DEC/HOPG, extracted from experimental 3D-AFM maps of DEC/flat HOPG interfaces. Inset is a schematic illustrating that $r_{sl}$ corresponds to the distance between a carbon atom (in HOPG) and the geometric center of a DEC molecule. **b**, SLS-calculated flat area density map, obtained by plugging in the atomic positions of four graphene layers (shown in the bottom, as model for HOPG basal plane) and the ETCF in **a** to Eq. 1. **c**, Liquid density variation ($\Delta\rho/\rho_{bulk}$) vs z curves of DEC/flat HOPG obtained from different methods. AFM-SLS curve (black) is obtained by averaging along the x direction in **b**. 3D-AFM experimental density curve (blue) is converted from the measured force curves (Supplementary Note 5). MD curve (red) is extracted by averaging along x in the MD-simulated x-z density map (Extended Data Fig. 5a).

**d–f,j–l**, Liquid density maps from AFM-SLS model (**d–f**) and MD simulation (**j–l**) near HOPG mono-step (**d,j**), bi-step (**e,k**), and tri-step (**f,l**). Gray dots in **d–f** and pink lines in **j–l** correspond to the HOPG substrate. Red dashed lines in **d–f,j–l** mark the boundaries of the transition region where the liquid layers are non-flat (processing method shown in Supplementary Fig. 10). Full 3D atomic structures of the constructed solids in **d–f** are shown in Supplementary Figs. 5 and 7. **g–i,m–o**, Horizontal density profiles along the liquid regions marked by the corresponding white dashed rectangles in **d–f,j–l**, respectively. The maximal $\Delta\rho/\rho_{bulk}$ value within the local z range is plotted at each x. Scale bars (**b,d–f,j–l**): 0.5 nm for both x and z.

At the planar basal plane area, AFM-SLS predicts flat, layered distribution of the interfacial liquid with a vertical periodicity of 4.0 Å (Fig. 2b), consistent with 3D-AFM imaging results (Fig. 1a). This spatial pattern directly reflects the superposition of ETCF over the substrate carbon atoms: 1) superposition from the dense, periodic packing of carbon atoms results in uniform in-plane density distribution of the liquid, since the intermolecular distance of DEC (4.0 Å) is much larger than the atomic bond length of HOPG (1.4 Å); and 2) the oscillating profile of $h_{eff}(\boldsymbol{r}_{sl})$ translates into the oscillation of liquid density along the z direction (with the same periodicity) after the linear superposition (see derivations in Supplementary Note 4).

To benchmark the AFM-SLS results, we further perform MD simulation of DEC/flat HOPG (system configuration shown in Supplementary Fig. 9a). We average the AFM-SLS x-z density map (Fig. 2b) over x to obtain the density vs z curve, and compare with results from 3D-AFM experiments and MD simulation (Fig. 2c). The conversion from 3D-AFM force to liquid density is discussed in Supplementary Note 5. We observe remarkable consistency in both the qualitative damped oscillation pattern and the quantitative density values among these three methods. These results verify the validity of the SLS method. MD simulation further reveals the anisotropic orientation of the interfacial liquids, particularly those in the first layer (Extended Data Fig. 5a,c,d,e). Despite the rich internal molecular structure and anisotropic orientation of the interfacial DEC molecules, the geometric center approach in the SLS model remains valid.

SLS density maps of the DEC at HOPG step edge areas (Fig. 2d–f) also closely resemble the experimental results (Fig. 1c–e), both revealing the same layer crossover and layer termination effects. Layer crossover is a manifestation of the mobile nature of liquid molecules, which rearrange their spatial distribution to minimize the overall free energy. This spatial reconfiguration likely involves a significant entropic contribution that disfavors conformal coverage at the atomic step sites. Near the atomic steps, we further observe a "transition region" (marked between the red dashed lines in Fig. 2d–f) where the liquid layers gradually change their z positions while remaining connected. The transition region occurs due to the competition of the carbon atoms' contribution to the liquid density through the ETCF, as the substrate at the left vs right side of the step edge favors different liquid layer positions. For larger step heights, the impact of the extra substrate layers at the left side of the step edge becomes more pronounced, causing the transition region to extend over a larger volume (wider angle between the boundary lines).

In addition to the z changes of layer positions, we further observe step edge-modulation of liquid density along the x direction (Fig. 2g–i). Near the atomic steps, liquid density profiles exhibit horizontal damped oscillations superposed on top of the average density of the corresponding layer. These oscillations have an average periodicity of $2\pi/\alpha_1 = 4.1 \pm 0.2$ Å and decay length of $1/\alpha_2 = 4.5 \pm 1.0$ Å (see Supplementary Table 2), close to those of the overall vertical density oscillations. We attribute the horizontal oscillation features to interference effects originating from the sharp edge termination and propagating along the x direction, again demonstrating the SLS principle.

To cross validate the AFM-SLS predictions on the step edge-induced liquid density variations, we perform MD simulations on the same DEC/HOPG atomic step systems (Supplementary Fig. 9b–d and Fig. 2j–o). The MD results reveal nearly identical layer crossover effects as the SLS predictions, including the triangular-shaped transition regions with similar size and boundary angles (Fig. 2j–l). At the transition region, both the density and molecular orientation of the DEC change abruptly, as a result of the layer crossover (Supplementary Fig. 4 and Extended Data Fig. 5b,f,g). In addition, we observe similar interference effects along the x axis near the step edges (Fig. 2m–o). The average periodicity and decay length of these oscillations are 4.2±0.1 Å and 3.4±0.6 Å, respectively (Supplementary Table 2), close to the AFM-SLS predictions. These quantitative agreements reveal the accuracy of SLS in predicting liquid density distributions at complex interfaces.

Note that horizontal liquid density oscillations have not been observed in experimental force maps. Additionally, the transition regions of the liquid layers near step edges are wider in the 3D-AFM x-z maps compared to those in AFM-SLS and MD results. These differences are likely due to AFM tip convolution effects that result in lower lateral resolution (Extended Data Fig. 3 and Supplementary Note 1).

**Interfacial liquid structure with more complex and larger solid morphologies**

A key advantage of the SLS model is the ability to quickly predict interfacial liquid structure adjacent to solid surfaces with arbitrary morphologies. To further test this capability, we extend the SLS simulation to another type of commonly observed heterogeneous surface structure of HOPG, buried step edges, where one or multiple layers of graphene covers an atomic step[33]. We construct a series of these buried step systems with different ramp angles of the top graphene layer (Fig. 3a and Supplementary Fig. 11). Plugging in the atomic positions of the constructed substrates and the already-obtained ETCF function between carbon atoms and DEC (Fig. 2a) to Eq. 1, we obtain the SLS-modeled liquid density distribution (Fig. 3a). At an angle of 90°, we observe layer crossover effect, similar to the results at exposed atomic steps. As the angle decreases, we find more conformal coverage of the liquid layers. At a ramp angle of 20° or lower, the crossover effect can no longer be observed and the liquid layers fully conform to the height change of the underlying substrate, due to the smoothness of the surface morphology. This conclusion is independent of the number of graphene layers covering the atomic step (Supplementary Fig. 12).

Experimentally, the ramp angle of buried step edges has large variations, and can reach as low as ~0.5° [33,34]. We have imaged a number of these structures, and observe conformal coverage of DEC above all of them, due to the low ramp angles (0.5°–9°, Fig. 3b,e and Supplementary Figs. 3 and 13). These results are consistent with the SLS predictions and further illustrate the key role of solid morphology in modulating the interfacial liquid structure.

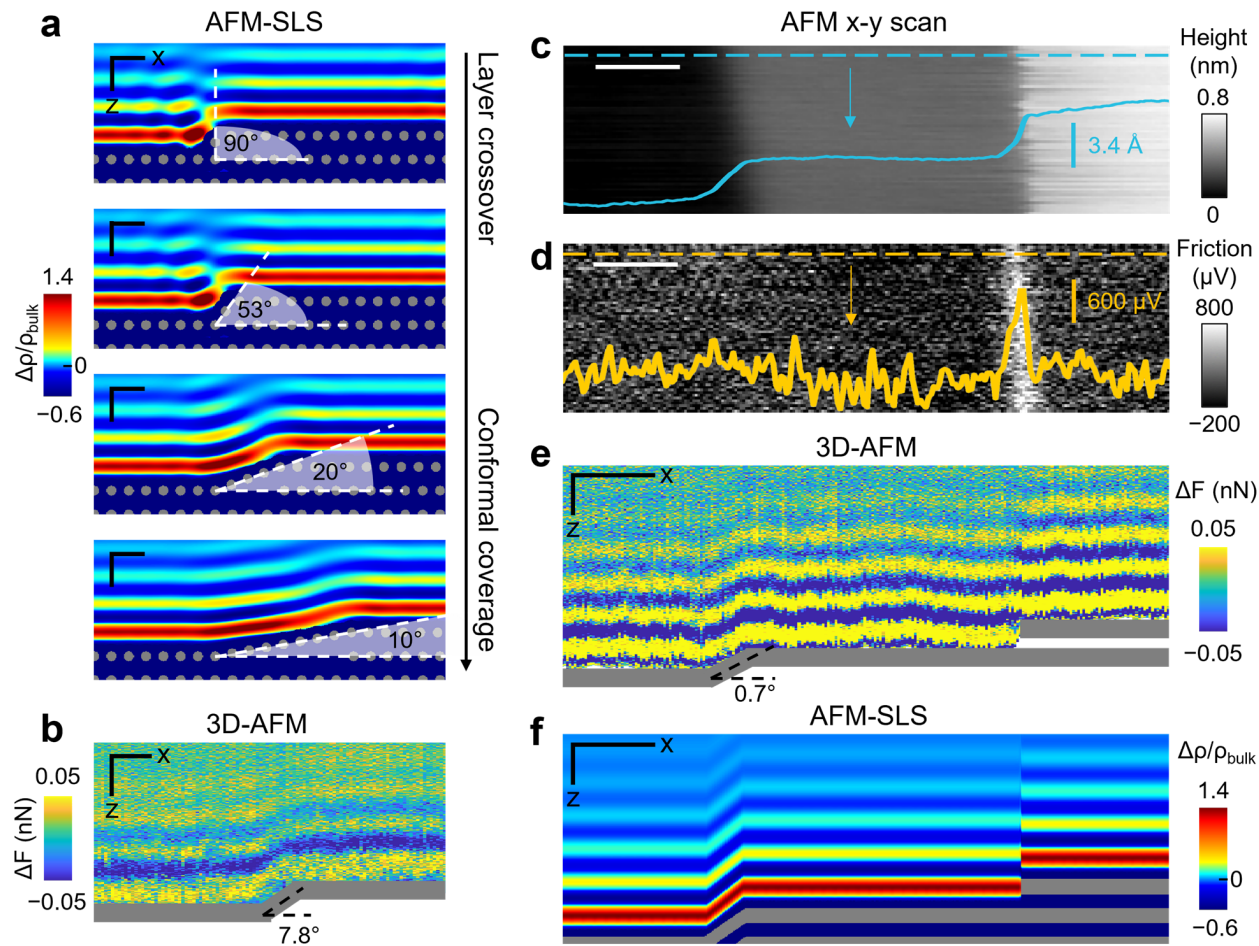

**Fig. 3 | Liquid structure at more complex and larger-scale DEC/HOPG interfaces. a**, AFM-SLS liquid density x-z maps near buried step edges of HOPG with different ramp angles. Full 3D atomic structures of the constructed solids are shown in Supplementary Fig. 11. **b**, A 3D-AFM x-z force map near a buried step edge of HOPG. **c,d**, DC mode AFM x-y images of height (**c**) and friction (**d**) of a 360 nm-wide region of HOPG immersed in DEC. The cross-sectional height and friction profiles extracted from the dashed line regions are overlaid on top of the x-y images. **e**, 3D-AFM (AC mode) x-z force map taken at the same region marked by the dashed lines in **c** and **d**. **f**, AFM-SLS x-z liquid density map where the solid structure is constructed according to **e**. Full 3D atomic structure of the constructed solid is shown in Supplementary Fig. 14. Scale bars: 0.5 nm for x and z (**a**), 2 nm for x and 0.5 nm for z (**b**), 50 nm (**c,d**), 50 nm for x and 0.5 nm for z (**e,f**).

The straightforward analytical nature of SLS also enables direct liquid structure calculations of interfacial systems at arbitrary size scales. This capability can be critical for understanding and predicting systems with a size of 100 nm and larger, such as whole living cells, microparticle electrodes in batteries, and ion exchange membranes[5,35,36]. As a demonstration, we perform 3D-AFM imaging of an HOPG region with a lateral length of 360 nm. As shown in Fig. 3c,d, we observe the co-existence of two heterogeneous sites, one exposed atomic step and the other buried, as confirmed from the x-y surface images of height and friction. Specifically, the exposed step has larger friction than the flat HOPG region, while buried step has nearly the same friction as the flat area (Fig. 3d), consistent with previous reports[33,34]. In addition, the height transition is sharper at the exposed step compared to the buried step (Fig. 3c). 3D-AFM x-z map in the same region reveals horizontal layered liquid structure at the flat regions, layer crossover at the exposed step edge, and conformal coverage at the buried step (Fig. 3e), all of which are reproduced by SLS simulation (Fig. 3f and Supplementary Figs. 14 and 15). Such size scales are highly challenging/demanding for all-atom MD simulations[5,35]; in contrast, SLS offers a direct solution for the interfacial liquid structure at arbitrarily large scales, with minimal computational cost.

**Interfacial structure of more complex liquids**

To further examine the generality of the SLS model, we investigate the interfacial structure of a series of more complex liquids, including ethylene carbonate (EC) and DEC mixtures with a series of molar ratios (commonly used in lithium-ion batteries[18]), 1 m (molal) lithium bis(trifluoromethanesulfonyl)imide (LiTFSI) in DEC (a battery electrolyte[37]) at different electrode potentials, pure water, and 0.01 M $K_2SO_4$ aqueous solution at different potentials (Fig. 4 and Extended Data Fig. 6). Potential is controlled using the electrochemical 3D-AFM method developed in our lab[2,21,38,39].

3D-AFM maps of these liquids at flat HOPG areas reveal similar damped oscillation patterns (Extended Data Fig. 6 and Supplementary Fig. 16), despite the distinct and highly complex compositions of the series of liquid systems. These oscillation features are consistent with previous 3D-AFM and X-ray scattering measurements at the interfaces between multi-component liquid electrolytes and flat substrates[3,9,21,40,41]. However, a quantitative understanding of these interfacial liquid structures and their extension to heterogeneous solid surfaces have been lacking.

Here we propose that Eq. 1 remains valid for multi-component liquid systems, where the ETCF can correspond to either one liquid component or the sum of all the components. If individual component's ETCF is used, Eq. 1 will produce the partial density distribution of the corresponding component. On the other hand, if the combined ETCF is plugged into Eq. 1, we obtain the total density of the liquid (sum of the partial density of every component). As 3D-AFM force is sensitive to the total density of liquid species[2,19–21], the ETCF extracted from 3D-AFM results will roughly correspond to the combined response of all the liquid components.

Since damped oscillations have been universally observed in the multi-component liquid systems that we have measured, we hypothesize that Eq. 2 is also valid for these interfacial liquids. However, in these complex liquids, 3D-AFM force curves tend to have large fluctuations at the first liquid peak (Extended Data Fig. 6), likely due to variations in the solvation condition at the end of the AFM probe, as previously reported[21,42]. As a result, it is highly challenging to quantify the first peak density of these complex liquids solely based on 3D-AFM data. To ensure reliable comparison of different liquid systems, here we set $A_1 = 0$ in Eq. 2 to simplify the ETCF formula. Supplementary Figs. 17 and 18 reveal that this simplification does not result in large deviation from the realistic interfacial liquid density profile. Combining the 3D-AFM maps at flat HOPG sites with Eq. 1 and the simplified Eq. 2, we obtain the ETCF for the series of complex liquid systems at carbon surfaces (Supplementary Note 6 and Supplementary Table 3). Since 3D-AFM data and simplified Eq. 2 are used, this ETCF extraction and subsequent SLS calculation protocol is referred to as "AFM(simp)-SLS". As shown in Fig. 4a and Extended Data Fig. 7, all ETCFs of the complex liquids exhibit damped oscillation patterns, similar to that of the pure DEC–carbon system.

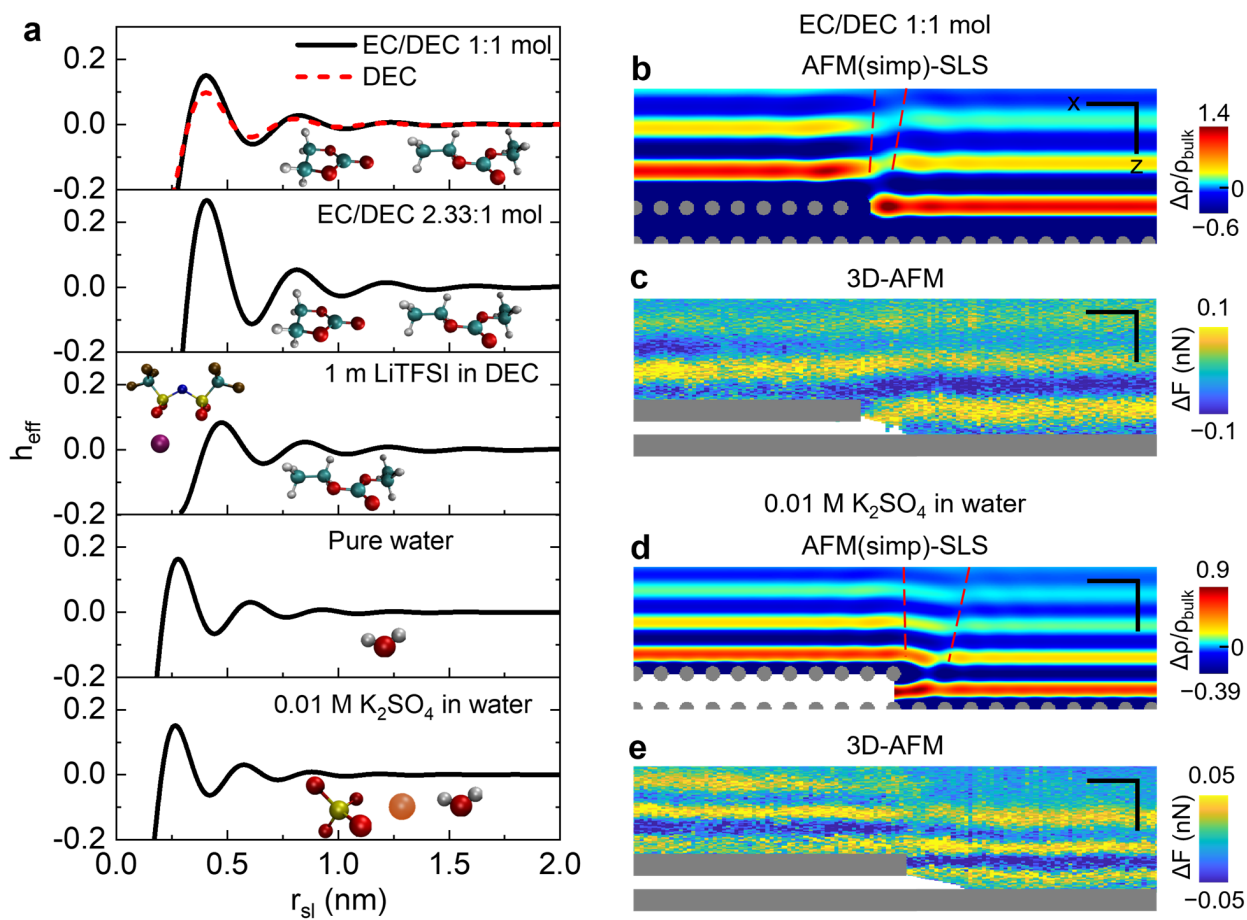

**Fig. 4 | Interfacial structure of more complex liquids. a**, ETCFs of a series of liquid systems at HOPG surface. ETCF is extracted using the AFM(simp)-SLS protocol (Supplementary Note 6) with parameters listed in Supplementary Tables 1 and 3. Molecular structures of the liquid components are also shown on the plots. Color code: carbon, green-cyan; oxygen, red: hydrogen, gray; sulfur, yellow; fluorine, brown; nitrogen, blue; lithium, purple; potassium, orange. **b–e**, x-z maps of molecular density from AFM(simp)-SLS (**b,d**) and x-z maps of 3D-AFM force (**c,e**) of EC/DEC mixture with 1:1 molar ratio (**b,c**) and 0.01 M $K_2SO_4$ in water (**d,e**) near HOPG mono-steps. Full 3D atomic structures of the constructed solids in **b,d** are shown in Supplementary Fig. 5. In all panels, the solvent-only systems are measured at open-circuit, 1 m LiTFSI in DEC at −0.5 V vs Pt, and 0.01 M $K_2SO_4$ at 1 V vs Ag/AgCl. Scale bars: 0.5 nm for x and z (**b,d**), 2 nm for x and 0.5 nm for z (**c,e**).

For all the EC/DEC mixtures and LiTFSI/DEC electrolyte (at different electrode potentials) at HOPG surface, their ETCFs exhibit a nearly identical periodicity ($2\pi/\alpha_1 \approx 4.1 \pm 0.2$ Å) and decay length $1/\alpha_2 \approx 4.5 \pm 1.2$ Å (Fig. 4a and Extended Data Fig. 7). Similarly, the pure water and aqueous solution (at varying electrode potentials) at HOPG surface also exhibit mostly identical periodicity and decay length in ETCF, with $2\pi/\alpha_1 = 3.1 \pm 0.2$ Å and $1/\alpha_2 \approx 3.9 \pm 1.5$ Å, respectively. This periodicity is the same as the known intermolecular distance of bulk water[20,43,44]. Note that variations in the exact conditions of each AFM probe have negligible impacts on the extracted $\alpha_1$ and $\alpha_2$ parameters (Supplementary Table 4).

For bulk liquids, OZ theory has predicted that the intermolecular TCF of liquid mixtures has the same periodicity and decay length as that of a pure liquid consisting of the dominant component[30]. This effect can be intuitively understood as the minority species permeating through the "structural framework" defined by the dominant species. Our results reveal that such bulk liquid behavior manifests itself in the interfacial density profile, following the SLS principle (Eq. 1). In EC/DEC mixtures, since pure EC and pure DEC have almost the same periodicity and decay length (~4 Å for both species[40,45,46]), it is natural to expect their mixture to exhibit the same metrics in the ETCF. For 1 m LiTFSI in DEC, the salt to solvent molar ratio is 0.12:1; in 0.01 M $K_2SO_4$ in water, the molar ratios are 0.002:1 ($K^+$ vs $H_2O$) and 0.001:1 ($SO_4^{2-}$ vs $H_2O$). Therefore, it is expected that the $\alpha_1$ and $\alpha_2$ parameters are the same as those of the corresponding pure solvent/solid interfaces.

While the $\alpha_1$ and $\alpha_2$ parameters tend to converge to the same values as those of the dominant species, other parameters in the ETCF can be more sensitive to the liquid composition changes and/or interfacial interactions. In EC/DEC mixtures, we find that the oscillation amplitude $A_0$ increases as the ratio of EC becomes higher (Fig. 4a and Extended Data Fig. 7). This is likely due to a stronger intermolecular interaction between EC molecules compared to DEC[24,46]; for the same reason, pure EC is a solid at room temperature while DEC is a liquid. Upon the addition of 1 m LiTFSI in DEC, we observe a broadening of the first liquid layer, which manifests as an overall phase shift in the extracted ETCF (Fig. 4a and Extended Data Figs. 6 and 7). This layer broadening / phase shift is possibly due to the accumulation of $TFSI^-$ in the first layer, as a result of their strong interaction with HOPG[47].

At different electrode potentials, we observe no significant changes in the ETCF for both the 1 m LiTFSI in DEC and 0.01 M $K_2SO_4$ in water solutions (Fig. 4a and Extended Data Figs. 6 and 7). This is consistent with our previous study of a series of aqueous solutions[21]. In the actual experimental systems, it is likely that the first liquid density peak will change at sufficiently large electrode potentials due to charge accumulation/depletion, although the first peak quantification by 3D-AFM is not always straightforward[21,42]. The upper layers are likely always dominated by the solvent regardless of the electrode potential, as the density of charges beyond the first liquid layer is oftentimes much lower than that of the solvent for the dilute solutions studied here[21].

We further examine the interfacial structure of these more complex liquids at heterogeneous substrate surfaces. We choose HOPG with exposed mono-step edge as the substrate (Supplementary Fig. 5), which has common well-defined structural features that enable the validation of the superposition principle. Using the ETCFs shown in Fig. 4a and Extended Data

Fig. 7m–o, we obtain SLS-modeled liquid density maps for EC/DEC mixtures (1:1, 1.82:1, and 2.33:1 mol) and 0.01 M $K_2SO_4$ in water (Fig. 4b,d, Extended Data Fig. 8, and Supplementary Fig. 19a,c). In addition, we conduct 3D-AFM measurements of the same interfacial systems, with results shown in Fig. 4c,e and Supplementary Fig. 19b,d. For all the systems, SLS predictions match well with the corresponding 3D-AFM maps, both revealing layer crossover effects. EC/DEC mixtures exhibit nearly identical spatial patterns near the atomic step compared to pure DEC/HOPG, due to their similar ETCFs. In the 0.01 M $K_2SO_4$ aqueous solution, liquid layers bend downwards when crossing from the left (above the upper HOPG surface) to the right (Fig. 4d,e), in contrast to the upward bending observed in pure DEC and EC/DEC mixture systems. This difference again manifests the superposition principle: a liquid layer on the left always connects to the nearest-neighbor layer on the right of the atomic step. Since the interlayer distance of aqueous solutions (~3.1 Å) is slightly smaller than that of the HOPG step height (3.4 Å), layer n (n=1, 2, 3, etc.) on the left needs to bend downwards when connecting to the nearby layer n+1 on the right; for the same reason, DEC and EC/DEC, with slightly larger interlayer distance (near 4.0 Å) than HOPG step height, exhibit upward bending. This nearby-layer-connection effect is also observed for EC/DEC mixtures adjacent to exposed bi-step edges (Supplementary Fig. 20).

In addition, SLS density maps of EC/DEC also show horizontal density oscillations next to the step edge (Extended Data Fig. 8b,d), with average periodicity of 4.3±0.3 Å and decay length of 3.6±0.1 Å, similar to those of pure DEC. The transition region boundary angles for all EC/DEC and pure DEC at HOPG mono-steps are also similar (~6°–9°). These similarities arise from the comparable ETCF periodicity and decay length characteristics of the EC/DEC and DEC systems. In contrast, changes in oscillation amplitude of the ETCF do not have a direct impact on the spatial pattern of the liquid distribution.

For the aqueous solution system, the horizontal density oscillation differs markedly from that of the EC/DEC systems. Even at the flat HOPG area, we observe in-plane oscillations in the first aqueous layer, with a periodicity of 2.5 Å, commensurate with the HOPG substrate (Extended Data Fig. 8e,f). This is because the first peak of the aqueous ETCF lies very close to the origin (2.7 Å, Fig. 4a), resulting in the partial projection of substrate atomic density onto the first liquid peak after applying the SLS model. Although this in-plane density oscillation of aqueous solutions on HOPG is beyond the sensitivity limit of 3D-AFM, previous studies observed this effect on mineral and molecular crystal surfaces with larger lattice constants using both 3D-AFM and MD simulation[26,48], further corroborating our SLS predictions. In upper liquid layers, due to the larger separation from the flat HOPG surface, the in-plane density becomes constant without oscillations. Near the HOPG mono-step site, we find the horizontal profile of the aqueous liquid density to be influenced by both the HOPG basal plane lattice and the edge termination (Extended Data Fig. 8g,h). Namely, the second aqueous layer on the right of the step edge exhibits damped oscillation with similar periodicity and decay length as the vertical oscillations (~3 Å), which resembles the behavior of (EC/)DEC (Fig. 2g–i and Extended Data Fig. 8b,d); in contrast, the first aqueous layer exhibits an initially larger oscillation period (~3.2 Å) with large amplitude next to the step edge, and gradually decays to the 2.5 Å periodicity with smaller amplitude after ~4 periods away from the step site.

Furthermore, the transition region boundary angle for the aqueous solution (12.2°, marked on Fig. 4d) is larger than that of the (EC/)DEC systems. This is likely due to the smaller mismatch between the liquid layer periodicity (0.31 nm) and the HOPG step height (0.34 nm), which produces a smoother structural transition and consequently a larger boundary angle.

These series of results confirm the capabilities of SLS to accurately predict liquid structures at most of the practically relevant solid–liquid interfaces, regardless of the solvent and solute species in the liquid electrolyte.

**Conclusions and outlook**

In summary, we have systematically investigated the structure of interfacial liquids at solid surfaces with a large range of complexity and size scale. Inspired by the experimentally observed real-space interfacial liquid configurations, we have developed the SLS model to interpret and predict liquid structures near arbitrary solid surfaces. This model is benchmarked against experimental results and MD simulations, and bridges the complexity gap in our existing understanding of interfacial liquids. SLS is broadly applicable in realistic solid–liquid interfaces, where the liquids include a diverse set of organic and aqueous solvents and electrolytes.

To further demonstrate the capabilities and generality of SLS, we perform more SLS modeling of many other systems. Extended Data Fig. 9 shows the SLS prediction of DEC distribution adjacent to other complex carbon substrates. A single-atom vacancy at the HOPG surface produces only a localized perturbation, manifested as a slight downward bending in the first liquid layer, while the surrounding liquid structure remains largely unchanged (Extended Data Fig. 9a). In contrast, a flat graphite edge plane induces liquid layering parallel to the substrate surface, with pronounced lateral density variations (Extended Data Fig. 9b,d). The periodicity of these lateral oscillations is 0.68 nm, the same as the substrate (vertical, ABAB-stacked graphene layers). For a rough edge plane, due to the strong substrate disorder, the adjacent liquid exhibits significantly disrupted ordering and rich interference patterns (Extended Data Fig. 9c,e).

Furthermore, SLS can be applicable beyond solid–liquid interfaces, in the general systems of fluids adjacent to rigid bodies. As demonstrations of this broader scope, we apply SLS to two systems reported in literature, liquid helium surrounding magnesium atoms[49] and 2D molecular gas near immobile sites[50] (Supplementary Note 7). As shown in Extended Data Fig. 10, both the spatial interference patterns and quantitative density profiles predicted by SLS are in good agreement with reported results produced from density functional theory (DFT), scanning tunneling microscopy (STM), and kinetic Monte Carlo simulations.

We have not encountered any solid–liquid interface system where SLS predictions strongly deviate from existing validated results. The exact scope of applications of SLS is a topic for future studies (Supplementary Note 8).

We have demonstrated the ease of integration of the SLS model with 3D-AFM experimental inputs, bridged by the ETCF descriptor. Following similar protocols, we have also successfully integrated MD inputs into SLS, and the resulting MD-SLS predictions are consistent with AFM-SLS (Supplementary Figs. 17 and 18). The straightforward analytical nature of SLS will also enable its integration with many other experimental and simulation methods, such as X-ray scattering and DFT, for multi-scale modeling of heterogeneous systems where solid–liquid interfaces are key components.

## Methods

### Probe and sample preparation

Ethylene carbonate (99%), diethyl carbonate (>=99%), and LiTFSI (99.95%) were all purchased from Sigma-Aldrich. Ultrapure deionized water (Milli-Q water) was obtained from the Synergy UV water purification system (MilliporeSigma). $K_2SO_4$ (anhydrous, >=99%) was purchased from Acros Organics. HOPG (ZYH grade, 12 mm × 12 mm × 2 mm) was purchased from Structure Probe Inc. AFM probes were purchased from NanoAndMore (PPP-NCHAuD) and Asylum Research (FS-1500AuD). The tips were made of silicon terminated with a native oxide layer. For each AFM probe we used, we measured its detailed mechanical, optical, and resonance metrics, and provided all of them in Supplementary Table 5. The cantilever was soaked in acetone for ~0.5 h, followed by soaking in isopropanol overnight (~12 h). Then, it was soaked in ultrapure deionized water for ~3 h before being cleaned by UV ozone (ProCleaner Plus system, BioForce Nanosciences) right before 3D-AFM measurement. After assembling the AFM liquid cell, HOPG was quickly cleaved using Scotch tape, and about 120 µL of liquid was dropped on the fresh graphite surface using pipette (Eppendorf North America). The whole AFM liquid cell was purged and then sealed in argon gas. A Pt ring was used as quasi-reference electrode for all experiments where an electrode potential was applied. In aqueous solutions, we further calibrated the Pt quasi-reference against Ag/AgCl, and found that 0 V vs Pt is equivalent to 0.22 V vs Ag/AgCl, as reported in our prior publication[21].

### AFM measurements

AC, amplitude-modulation mode 3D-AFM measurements were performed using a Cypher ES atomic force microscope (Asylum Research, Oxford Instruments). Thermal tuning and cantilever tuning were carried out in air and liquid to determine the spring constant, quality factor, inverse optical lever sensitivity (InvOLS), and resonance frequency. Prior to and right after acquiring 3D maps, topographic images of the HOPG surface in the x-y plane were recorded to assess the surface cleanliness. 3D-AFM measurements were carried out in clean regions, where the surface features include flat terraces, exposed atomic steps, and/or buried step edges. During 3D-AFM measurements, the AFM tip scanned a three-dimensional volume above the surface, with a sinusoidal z motion at a rate of 10 Hz. AFM imaging parameters are provided in Supplementary Table 6. Typical scan volumes for flat surfaces, step edges, and larger-scale structures (e.g., those containing multiple step edges) were $10 × 10 × 3$ nm$^3$, $40 × 2.5 × 3$ nm$^3$, and $400 × 2.5 × 3$ nm$^3$ (x × y × z), respectively. Corresponding pixel resolutions were $64 × 64 × 25,000$ for flat surfaces and $256 × 16 × 25,000$ for areas containing step edges.

3D-AFM data were acquired sequentially as x-z frames, with y direction as the slow scan axis. For each x-z frame, four datasets were recorded: trace and retrace (forward and backward lateral motion along x) and approach and retract (downward and upward motion along z). These datasets yielded nearly identical maps and were used to verify data reproducibility; the same protocol and reproducibility examination were reported in our prior publication[2]. At each pixel, the spatial position (x, y, z) as well as the cantilever oscillation amplitude and phase were recorded.

For lateral deflection measurements, DC, contact mode scans were performed. Two datasets were recorded: trace and retrace. The friction map in Fig. 3d was obtained by subtracting retrace lateral deflection map from trace lateral deflection map (Supplementary Fig. 21). The latera force map in Extended Data Fig. 1h corresponds to the trace scan. Supplementary Fig. 13a shows a lateral force map obtained along the retrace scan.

All the AFM measurements reported in this work were conducted at room temperature.

### 3D-AFM data analysis

In AC mode 3D-AFM, each data point consists of x, y, scanner extension (Ext), amplitude, and phase. Tip–sample distance $d$ was calculated as $d = -Ext - amplitude + C$, where $C$ is a constant applied so that $d$=0 corresponds to the substrate surface. The measured amplitude and phase data were subsequently converted to force using the method initially described by Payam *et al*[51]. Both the phase–distance and force–distance curves contain two components, a long-range repulsive background and a short-range oscillatory component. Only the short-range component is directly related to the local liquid density variations[19]. To remove the repulsive background, we performed double-exponential fit and subtracted the fitted function from the original phase–distance and force–distance curves. In this work, the first minimum of the background-subtracted force ($\Delta F$) curve was used as the origin ($z = 0$) for plotting the z axis. If not otherwise noted, all the 3D-AFM x-z maps and z curves in this article show the background-subtracted phase or force.

Supplementary Fig. 22 shows an example of the force reconstruction and background subtraction process.

### SLS calculations

All SLS density maps were computed using MATLAB. Graphite substrates with hexagonal lattice (honeycomb structure) and ABAB stacking were constructed, with a carbon–carbon bond length of 0.142 nm and interlayer distance of 0.34 nm. The flat basal planes and exposed step edges used for SLS modeling are illustrated in Fig. 2b and Supplementary Figs. 5–7. Unless otherwise specified, exposed step edges in the SLS model adopt an armchair configuration. The choice of step type (armchair vs zigzag) does not have a measurable impact on the resulting liquid density maps, as demonstrated in Supplementary Fig. 23. For buried step edges and large-scale simulations (Fig. 3a,f), the exact substrate geometries are shown in Supplementary Figs. 11 and 14. For each SLS density map, at least ~3000 substrate atoms surrounding the target interfacial liquid region were included and distributed in all spatial directions (x, y, and z) (see the example in Fig. 2b), enabling accurate calculation of the liquid density throughout the target region using the superposition formula (Eq. 1). Each x-z liquid density map consists of a 500 × 500 pixel grid. At each pixel, the distance between the pixel location and every substrate atom was calculated as $r_{sl}$. The liquid density variation ($\Delta\rho(r_l)/\rho_{bulk}$) at a given pixel was obtained by summing $h_{eff}(r_{sl})$ over all substrate atoms (at $r_s$) following Eq. 1. Repeating this procedure over the entire grid produces the full SLS density map.

In typical calculations for flat and step-edge regions, approximately 3,000–10,000 substrate atoms were included. For larger-scale simulations, more substrate atoms were constructed. For example, the graphite substrate shown in Fig. 3f and Supplementary Fig. 14 contains 131,328 carbon atoms.

**Molecular dynamics simulations**

MD simulations of DEC solvent adjacent to HOPG flat area and different atomic steps (mono-step, bi-step and tri-step) were performed using the LAMMPS package[52]. The setups of HOPG/DEC/HOPG sandwich systems are shown in Supplementary Fig. 9. Each MD system includes two HOPG substrates, with each substrate constructed using multiple layers of graphene sheets (ABAB stacking). The interlayer distance of graphene is 0.335 nm. The flat HOPG system contains 4 graphene layers with a surface area of 5.97 nm × 6.24 nm. The HOPG containing step edges has 3 bottom graphene layers with an area of 9.95 nm × 6.24 nm, in addition to 1–3 upper layers with an area of 4.91 nm × 6.24 nm. The step edges have an armchair structure. The separation of the two HOPG/DEC interfaces within each simulated system is at least 12 nm everywhere to ensure DEC molecules in the channel center exhibit bulk-like behavior. The force field parameters of HOPG were taken from Caleman et al[53]. DEC molecules were modeled using the Automated Topology Builder[54–56]. Non-electrostatic interactions, including van der Waals (Lennard–Jones) interactions, were evaluated using a truncated cutoff scheme consistent with the employed force fields (Caleman et al. and the Automated Topology Builder). The long-range electrostatic interactions were computed using the particle mesh Ewald (PME) method[57]. A cut-off distance of 1.2 nm was used to compute both the electrostatic and non-electrostatic interactions. We applied periodic boundary conditions in the x and y directions. The system was non-periodic in the z direction. The simulations were performed under canonical (NVT) ensemble and the temperature was maintained at 300 K by using the Nosè–Hoover thermostat[58,59]. We applied a time step of 1 fs. The simulations were first equilibrated for 5 ns, followed by a production run for 35 ns. The MD trajectories of atomic coordination were saved every 1 ps. Each simulation was performed three times using independent initial configurations to generate statistically independent trajectories. The resulting observables were averaged to obtain the reported results.

**Methods for extracting horizontal density/force profiles**

To obtain horizontal density/force profiles for a segment of x-z map ($x_1 \leq x \leq x_2$, $z_1 \leq z \leq z_2$, where $z_1$ and $z_2$ can vary at different x), we first extracted the maximum density/force value within the analyzed z interval ($z_1 \leq z \leq z_2$) at each x. For MD and SLS results, the maximal liquid density was directly plotted as a function of x (from $x_1$ to $x_2$). For 3D-AFM force maps, the extracted maximal force values were further smoothened using MATLAB's "smoothdata" function with the "Gaussian" method and a window size of five adjacent points along the x direction. These smoothened maximal force values were plotted as a function of x (from $x_1$ to $x_2$).

**Data availability**

All data are included in the main article, Extended Data, and Supplementary Information. Source data will be provided prior to publication.

**Code availability**

SLS codes used to produce Fig. 2b are provided as a Supplementary Information file. To produce all other SLS maps in this article, the same codes were used, except that the substrate atomic positions (provided in Supplementary Figs. 5–7, 11 and 14) and/or ETCF parameters (specified in Supplementary Tables 1 and 3) were substituted.

**Acknowledgement**

Q.A., L.K.S.B., K.S.P., S.Z., F.Z., Y.L. and Y.Z. acknowledge support from the National Science Foundation under Grant No. 2339175, the Beckman Young Investigator Award provided by the Arnold and Mabel Beckman Foundation, and the Sloan Research Fellowship from the Alfred P. Sloan Foundation. Experiments were carried out in part in the Carl R. Woese Institute for Genomic Biology and in the Materials Research Laboratory (MRL) Central Research Facilities at the University of Illinois. We are grateful to Minjiang Zhu for discussion on the SLS model.

**Author contributions**

Q.A. and Y.Z. conceived the overall idea. Q.A., L.K.S.B., K.S.P., S.Z., and F.Z. conducted the experiments, guided by Y.Z. Q.A. performed the experimental data analysis. Q.A. and Y.Z. conceptualized the SLS model and performed calculations, with help from Y.L. and K.S.S. H.W. and N.R.A. performed MD simulations. Q.A. and Y.Z. wrote the manuscript, with help from all other authors. Y.Z. and N.A. supervised the work.

**Competing interests**

The authors declare no competing interests.

**Additional information**

**Supplementary information:** The online version contains supplementary material (a pdf file for Supplementary Notes, Figures, and Tables, and a zip file for SLS source code).

**Correspondence and requests for materials** should be addressed to Yingjie Zhang.

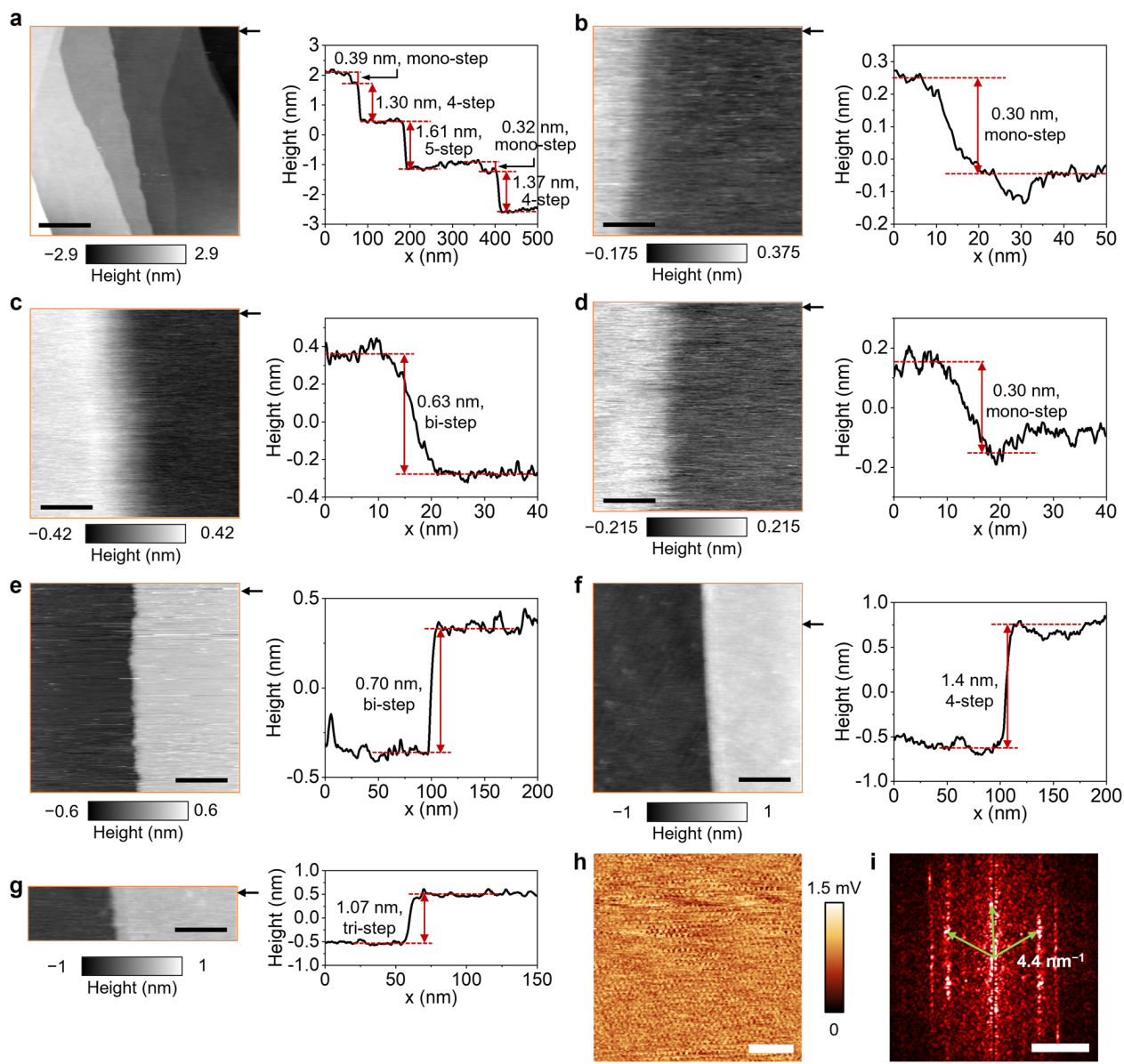

**Extended Data Fig. 1 | AFM x-y images of HOPG surface immersed in liquid.** Liquid systems and electrode potentials are: DEC at open-circuit (**a**–**d**), 0.01 M $K_2SO_4$ in water at 0 V vs Ag/AgCl (**e**), EC/DEC 1:1 mol (**f**) and 2.33:1 mol (**g**) mixtures at open-circuit, and 1 m LiTFSI in DEC at 1 V vs Pt (**h**,**i**). **a**–**g** are AC mode surface height maps, while **h** is a DC mode lateral force image. The height profiles at positions marked by the arrow (next to the left panels in **a**–**g**) are extracted and shown as the right panels in **a**–**g**. Across each step edge, the height change usually occurs over a width of several nanometers, instead of an infinitely sharp jump, due to the finite size of the AFM tip (see Extended Data Fig. 3). Scale bars: 125 nm (**a**), 12.5 nm (**b**), 10 nm (**c**,**d**), 50 nm (**e**,**f**), 37.5 nm (**g**), 2 nm (**h**), 5 $nm^{-1}$ (**i**). **h** resolves the atomic lattice of the HOPG surface, with the Fourier transform shown in **i**. From **h**,**i**, the HOPG lattice constant is extracted as ~0.23 nm, consistent with our previous observations and the known standard value[21,38,39].

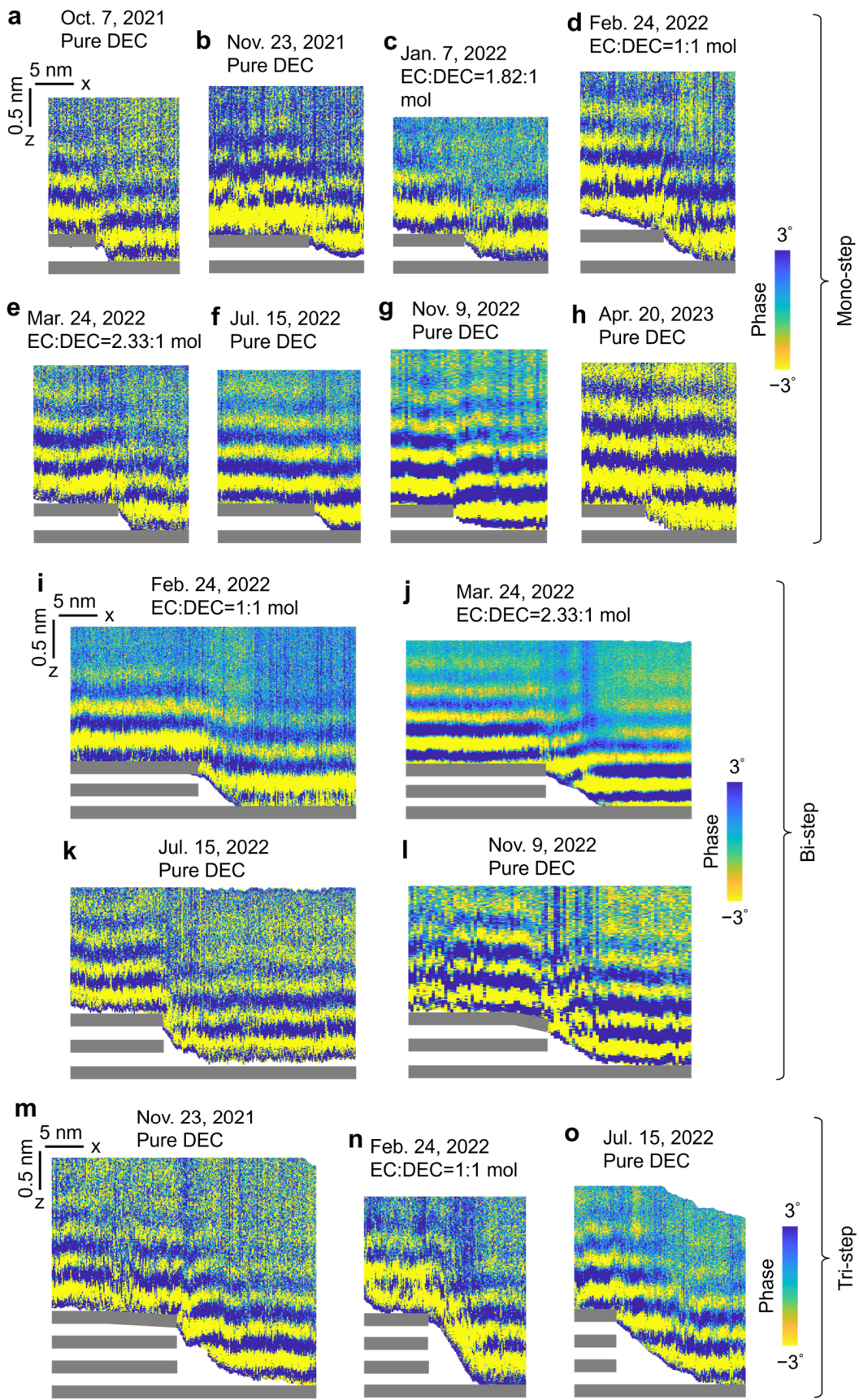

**Extended Data Fig. 2 | Gallery of 3D-AFM x-z phase maps of pure DEC and EC/DEC near exposed atomic steps of HOPG.** The measurement date and liquid system are specified on top of each map. All results are measured at open-circuit. These maps are obtained from a total of 8 different experiments with multiple types of AFM tips, measured at three types of exposed atomic steps (mono-step, bi-step, and tri-step). The measured tip parameters are provided in Supplementary Table 5. The gray lines at the bottom of each map correspond to the assigned position and structure of HOPG substrate. Low phase (yellow) regions in the map roughly correspond to the liquid layers. Layer crossover effect is universally observed in all the maps.

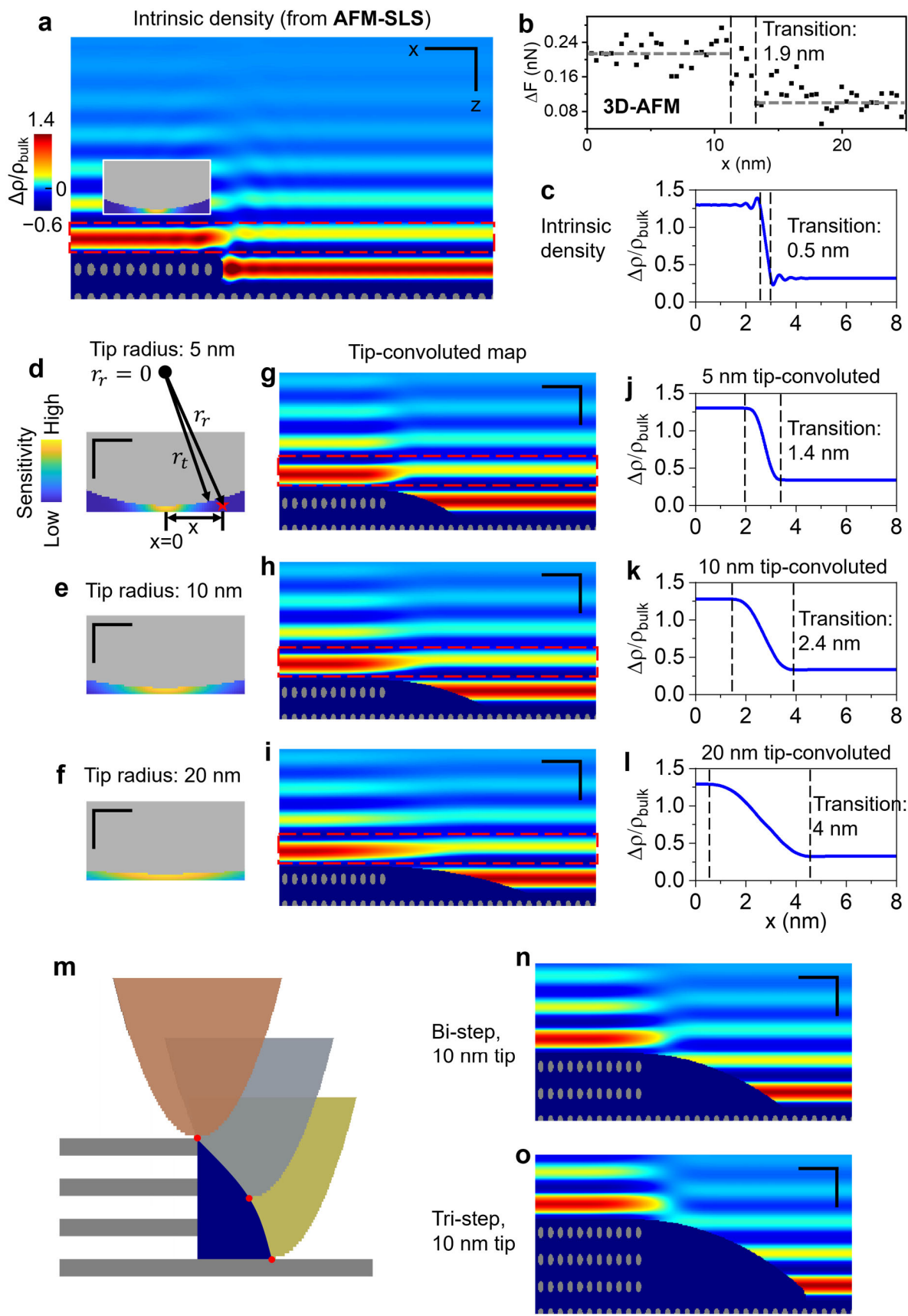

**Extended Data Fig. 3 | Tip convolution effect.** Simulated interfacial system: DEC/HOPG. **a**, Intrinsic liquid density map calculated by AFM-SLS. The inset on the left shows a simulated AFM tip that senses an average signal surrounding the tip apex (tip radius: 5 nm). **b**, Force vs x along a connected liquid layer across a step edge, obtained from an experimentally measured force map of DEC/HOPG mono-step (same as Supplementary Fig. 4d). The transition region is ~1.9 nm wide. **c**, Horizontal maximal density curve extracted from the red dashed box in **a**. The transition region is ~0.5 nm wide, narrower than that obtained from the corresponding experimental system shown in **b**. **d**–**f**, Simulated sensitivity maps surrounding AFM tips with a radius of 5 nm (**d**), 10 nm (**e**), and 20 nm (**f**). Regions with higher sensitivity have higher weighted contribution to the corresponding convoluted density map. **g**–**i**, Simulated tip-convoluted x-z density maps after dilation operation using the intrinsic density map in **a** and the corresponding sensitivity maps in **d**–**f**, respectively. **j**–**l**, Horizontal maximal density profiles extracted from the red dashed boxes in **g**–**i**, respectively. The transition region width increases with tip radius. **m**, Schematic illustrating the formation of an inaccessible area near the step edge due to the finite size of the tip. **n**,**o**, Simulated tip-convoluted x-z density maps illustrating the evolution of the inaccessible area as the step structure changes, including bi-step (**n**) and tri-step (**o**) (compare to the mono-step in **h**). Full 3D atomic structures of the constructed solids in **a**,**g**–**i**,**n**,**o** are shown in Supplementary Figs. 5 and 7. Scale bars: 1 nm for x and 0.5 nm for z (**a**,**g**–**i**,**n**,**o**), 0.5 nm for x and 0.25 nm for z (**d**–**f**). Color bars for **g**–**i**,**n**,**o** are the same as that shown on the left of **a**.

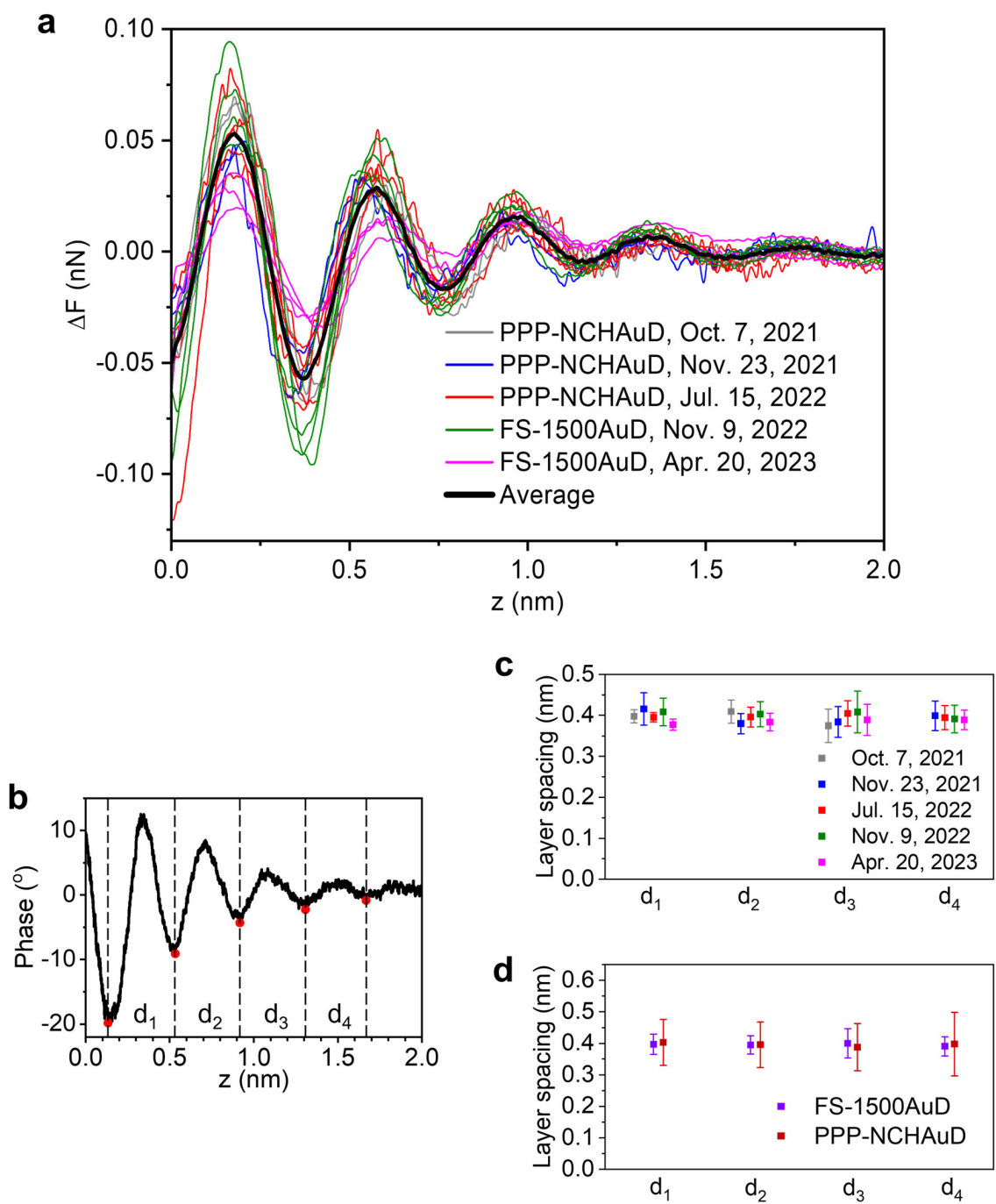

**Extended Data Fig. 4 | Reproducibility of 3D-AFM measurements.** Results are summarized from 3D-AFM force curves of DEC at HOPG flat areas at open-circuit. **a**, Force curves from 5 different experiments (measurement dates labeled) using two types of cantilevers (PPP-NCHAuD and FS-1500AuD). Each colored curve is an average of at least 150 single force curves. Black curve is the average of all colored curves. Similar force curves are observed in all five measurements. For each measurement, we further extract the characteristic parameters of the force curves following part of the AFM(simp)-SLS protocol, with results summarized in Supplementary Table 4. Similar parameters are obtained regardless of the measurement date or cantilever type,

further confirming the quantitative reproducibility of our 3D-AFM results. **b**, An example to illustrate the automated protocol for determining layer spacings. Local minima, identified using MATLAB's "islocalmin" function, are marked by red dots. The layer spacings of interfacial liquid are calculated as the distances between adjacent minima. $d_1$ is the distance between the first and second layer, $d_2$ the distance between the second and the third layer, and so forth. **c**, DEC layer spacings extracted from measurements at different dates. The spacings remain consistent among all experiments conducted over multiple years. Each data point (average value and standard deviation) represents calculation from over 128,000 individual phase curves. **d**, DEC layer spacings (average value and standard deviation) measured using different types of cantilevers. The spacings are found to be independent of the cantilever type. Over 600,000 individual phase curves are obtained and used for analysis using PPP-NCHAuD tips, and more than 300,000 curves using FS-1500AuD tips.

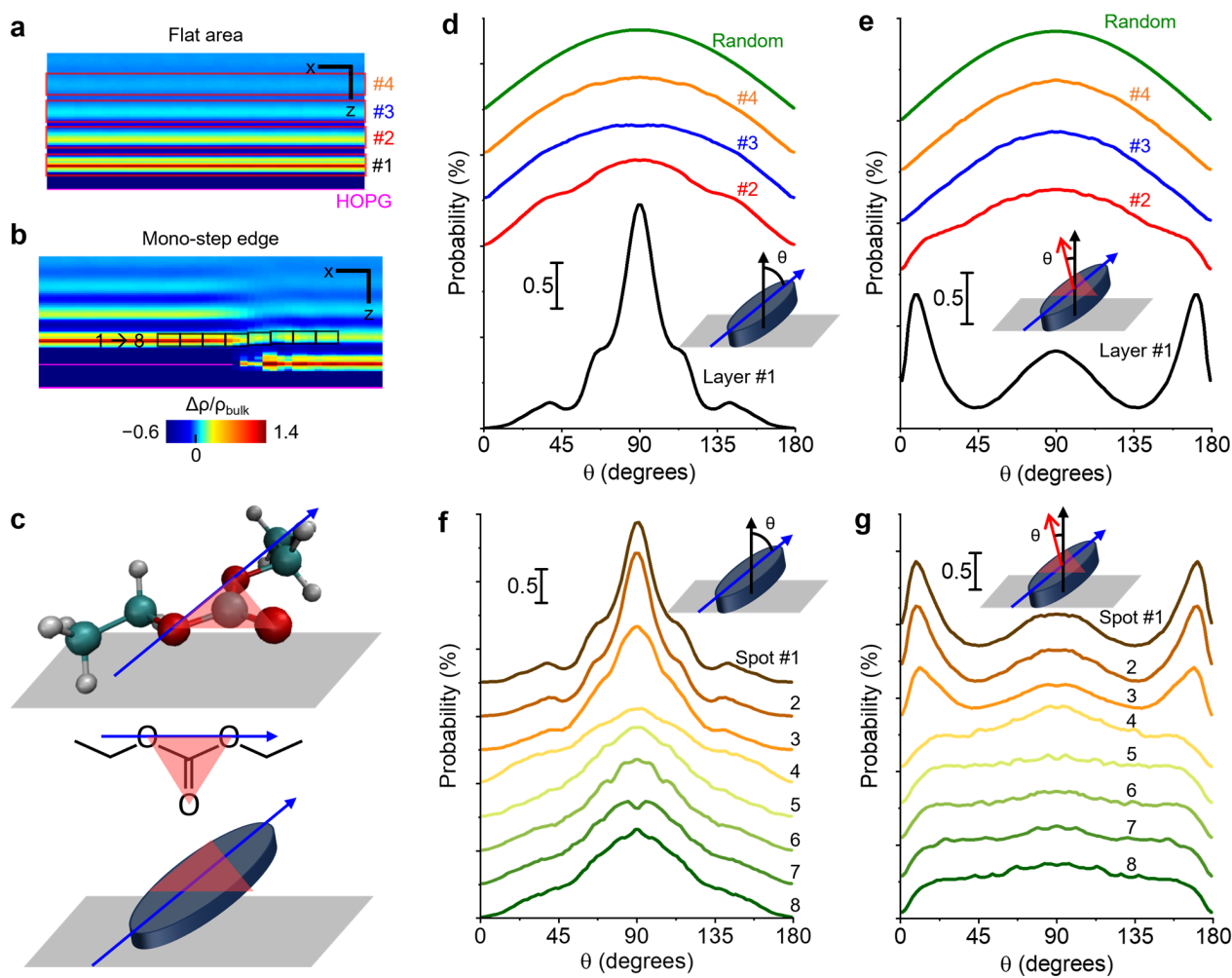

**Extended Data Fig. 5 | MD simulated molecular orientation at the flat and mono-step transition regions.** System: DEC/HOPG at open-circuit. **a**, MD simulated x-z density map of DEC at flat HOPG surface. Molecular orientation is calculated at the four high density regions marked by red boxes. **b**, MD simulated x-z density map of DEC near an HOPG mono-step edge (same as Fig. 2j). Molecular orientation is calculated at the 8 boxed areas across the layer crossover region. Scale bars (**a**,**b**): 0.5 nm for both x and z. **c**, Schematic of a DEC molecule and its simplified form. Green-cyan: carbon; red: oxygen; gray: hydrogen. The molecule is simplified as an elongated plate for the purpose of marking the angular orientations. The blue arrow marks the direction connecting the two oxygen atoms with single carbon–oxygen bond. The red triangle marks the plane formed by the three oxygen atoms. **d**,**e**, Simulated orientation distribution from the corresponding liquid layers marked in **a**. The angle is between the normal direction of the substrate and (**d**) the blue line connecting two oxygen atoms with single carbon–oxygen bond or (**e**) the normal direction of the red plane of three oxygen atoms in DEC. In the first liquid layer, DEC molecules are mostly parallel to the substrate. In layer 2, the molecules are largely randomly distributed with a slight tendency toward the flat configuration. In layer 3 and above, the molecular orientation is almost fully random. **f**,**g**, Orientation distribution from the eight spots marked in **b**. Molecules gradually evolve from a typical layer #1 orientation to a typical layer #2 orientation from left to right as layer crossover occurs.

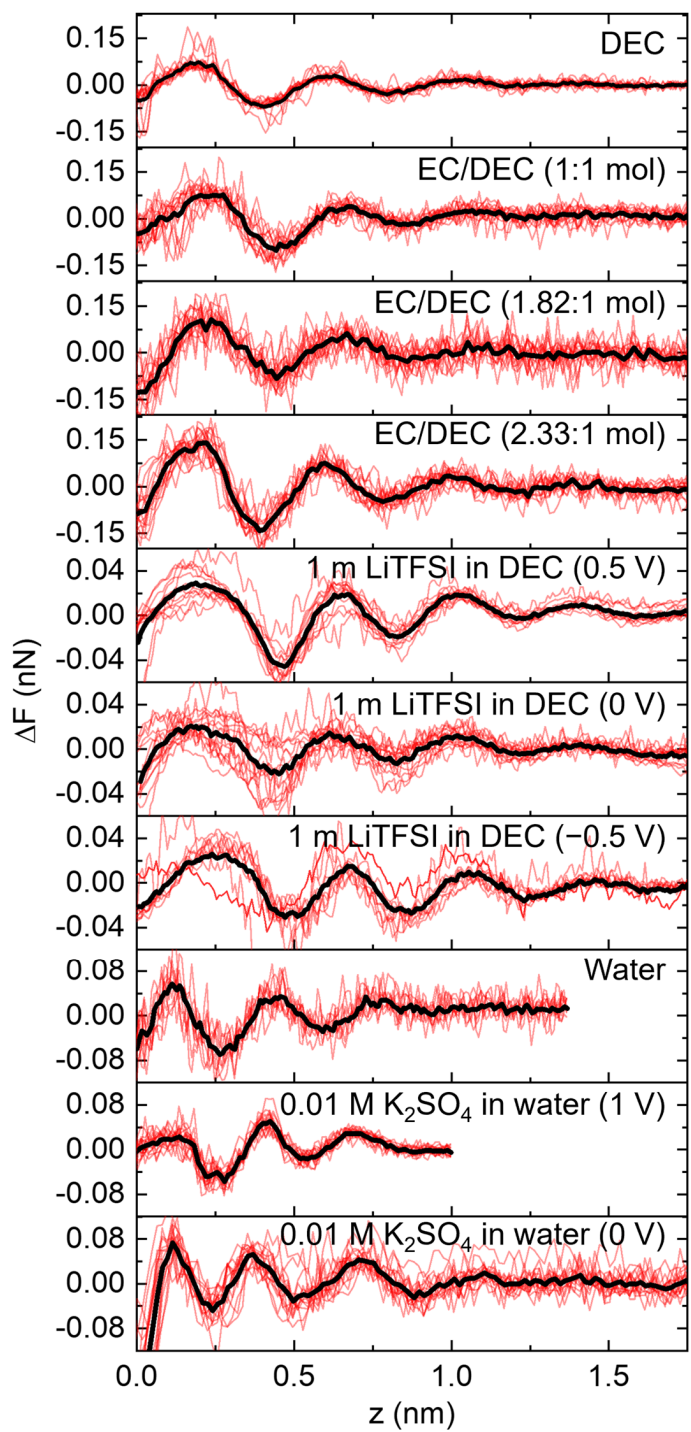

**Extended Data Fig. 6 | 3D-AFM force curves of more complex liquids at flat HOPG substrates.** Red curves are individual force vs z curves. Black curves are averaged from red curves. The solvent-only systems are measured at open-circuit. For liquid electrolytes, the marked potentials are referenced vs Pt for LiTFSI in DEC, and vs Ag/AgCl for 0.01 M $K_2SO_4$ in water. In our previous work, we conducted 3D-AFM measurements of 0.01 M $K_2SO_4$ in water on HOPG at a series of negative potentials vs Ag/AgCl, and observed no measurable differences in the force curves[21].

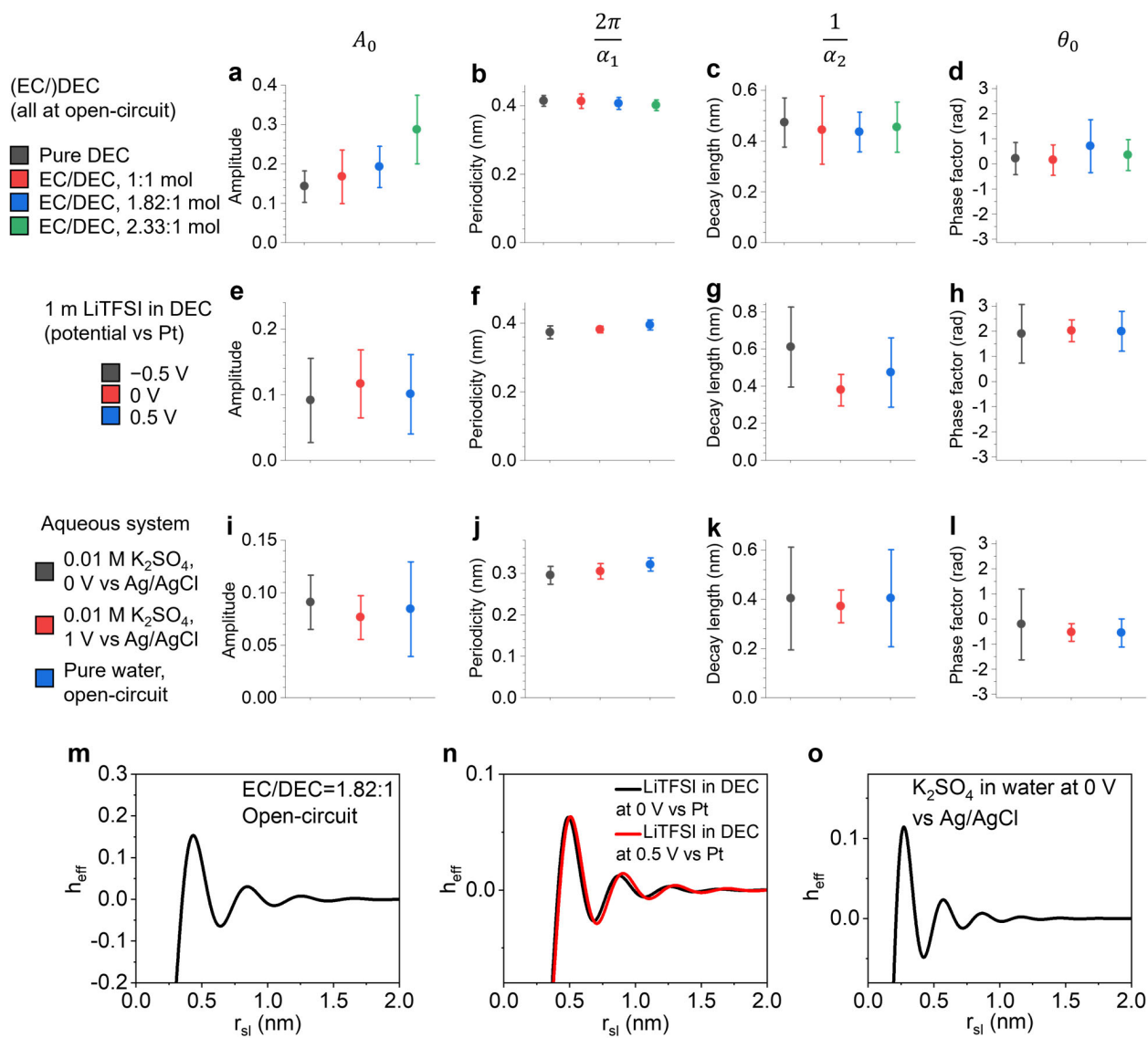

**Extended Data Fig. 7 | Additional ETCF quantification of more complex liquids at HOPG surface.** ETCF is extracted using the AFM(simp)-SLS protocol. **a–l**, Extracted ETCF parameters for pure DEC and EC/DEC mixtures at open-circuit (**a–d**), 1 m LiTFSI in DEC at different potentials (**e–h**), and water at open-circuit and aqueous solution at different potentials (**i–l**). The plotted parameters include oscillation amplitude $A_0$ (**a,e,i**), periodicity $2\pi/\alpha_1$ (**b,f,j**), decay length $1/\alpha_2$ (**c,g,k**), and phase factor $\theta_0$ (**d,h,l**). Each plotted average value and error bar (standard deviation) are obtained from a set of extracted ETCF functions, where each ETCF is obtained from fitting of a single force curve. Statistics for the (EC/)DEC system are extracted from more than one million force curves, whereas statistics for 1 m LiTFSI in DEC and for the aqueous system are each extracted from more than 82,000 force curves. **m–o**, ETCF plots for EC/DEC with 1.82:1 molar ratio at open-circuit (**m**), 1 m LiTFSI in DEC at different potentials (**n**) and 0.01 M $K_2SO_4$ in water at 0 V vs Ag/AgCl (**o**) constructed using the average parameters from **a–l**.

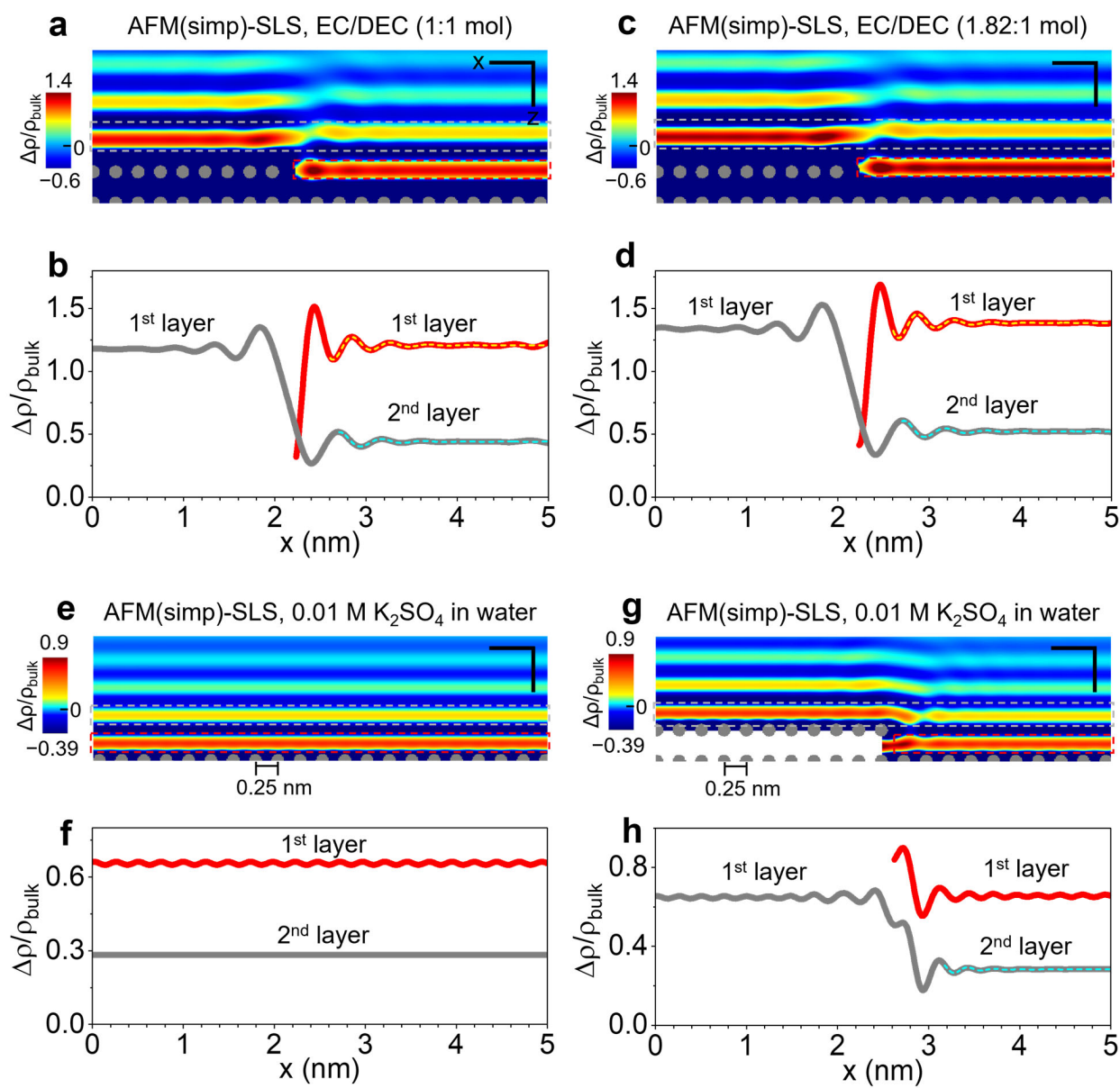

**Extended Data Fig. 8 | Horizontal density profiles of complex liquids at HOPG surfaces (flat or mono-atomic step region). a,c,e,g**, AFM(simp)-SLS x-z density maps for EC/DEC mixture (1:1 mol) at mono-step (**a**), EC/DEC mixture (1.82:1 mol) at mono-step (**c**), 0.01 M $K_2SO_4$ in water at flat area (**e**), and 0.01 M $K_2SO_4$ in water at mono-step regions (**g**). **a,c** are simulated for open-circuit and **e,g** are simulated for 1 V vs Ag/AgCl. Panels **a** and **g** are identical to Fig. 4b,d, respectively. Scale bars: 0.5 nm for x and z (**a,c,e,g**). **b,d,f,h**, Maximal density vs x profiles extracted from areas marked by the dashed boxes in **a,c,e,g**, respectively. Red density curves correspond to the red dashed boxes, while gray curves to gray dashed boxes. Yellow and cyan dashed curves are fits for red and gray curves, respectively. Fitting parameters are provided in Supplementary Table 2. Full 3D atomic structures of the constructed solids are shown in Supplementary Fig. 5 (**a,c,g**) and Fig. 2b (**e**).

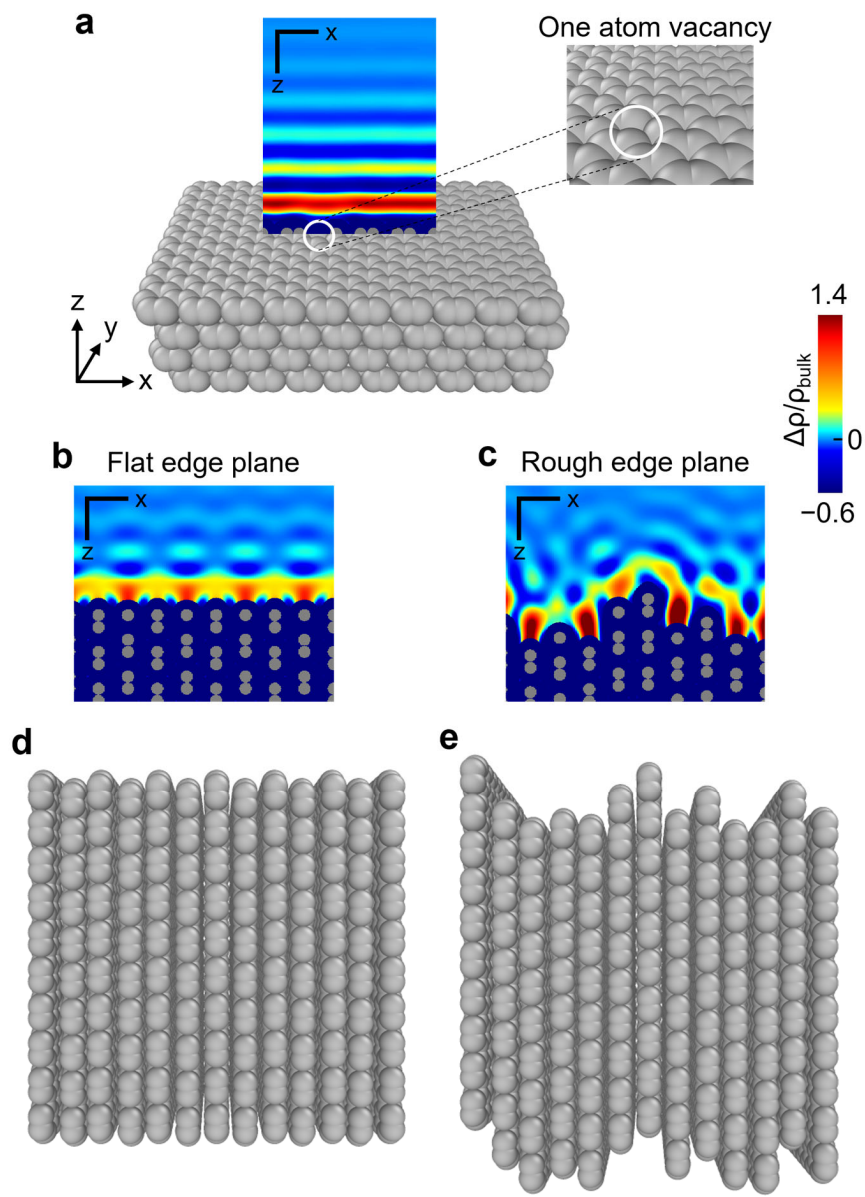

**Extended Data Fig. 9 | AFM-SLS maps of DEC adjacent to other complex carbon substrates.**
**a**, x-z liquid density map near HOPG containing a surface atomic vacancy. The white circle shows the position of the missing atom. **b,c**, x-z density maps of DEC adjacent to a flat edge plane (**b**) and a rough edge plane (**c**) of graphite. **d,e**, Schematics of the substrate configurations used in **b,c**, respectively, where the top substrate surface has zigzag edge structure (Supplementary Fig. 6). Scale bars: 0.5 nm for x and z (**a**–**c**).

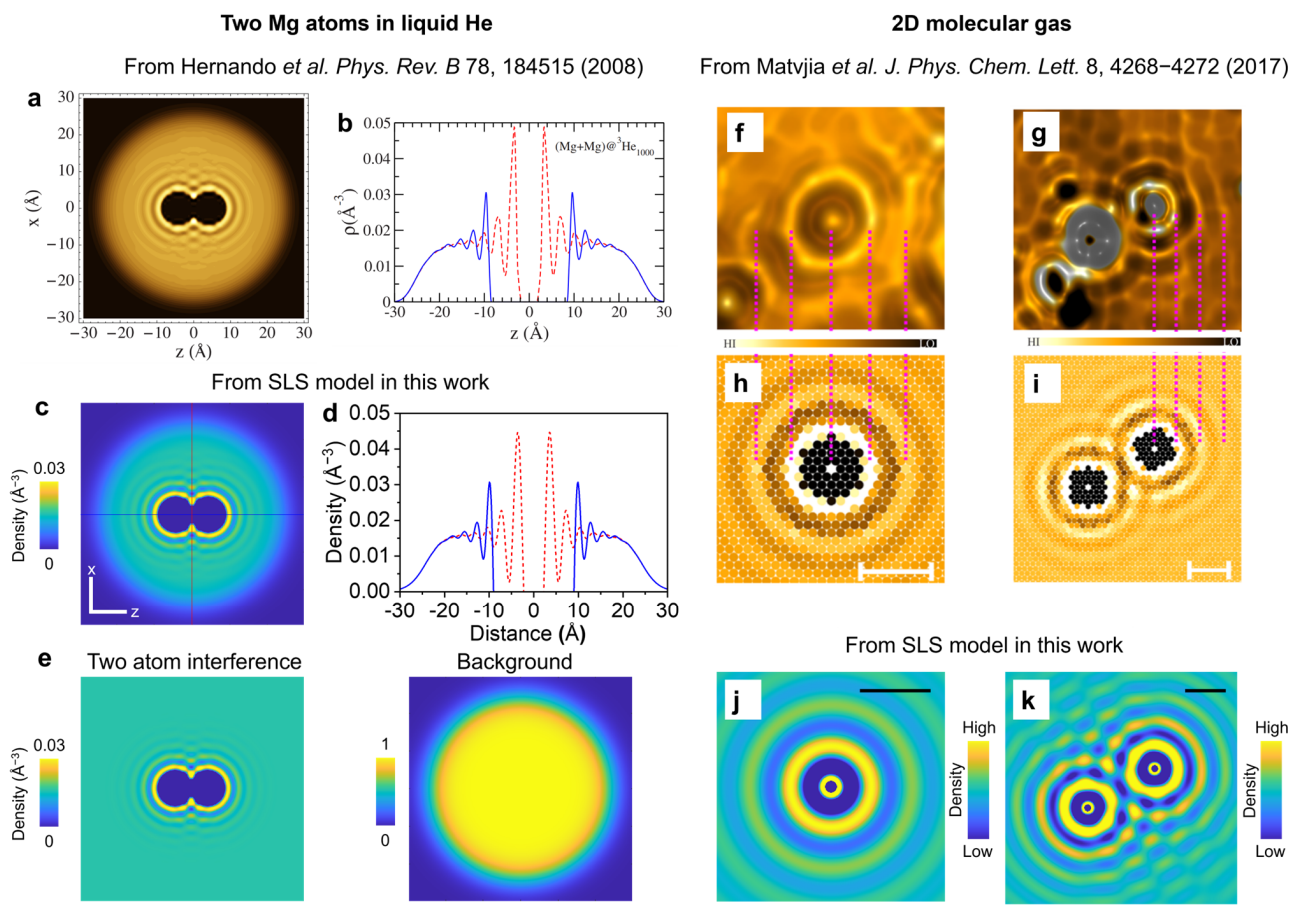

**Extended Data Fig. 10 | Application of the SLS model to reported systems beyond solid–liquid interfaces.** Panels **a** and **b** are DFT calculation results from Ref. [49]. Copyright 2008 American Physical Society. **a**, Map of helium density surrounding two Mg atoms in a liquid helium droplet. **b**, Helium density vs z at x=0 (blue solid line) and helium density vs x at z=0 (red dashed line). **c**, Helium density map produced from our SLS model with the same Mg atomic pair as in **a**. Scale bars: 1 nm for both x and z. **d**, SLS-modeled liquid helium density profile vs z at x=0 (blue solid line) and density vs x at z=0 (red dashed line). **e**, The interference pattern (left) and reflected sigmoid background (right) of **c**. They are multiplied to mimic the helium droplet boundary condition as reported in Ref. [49], producing **c**. **f**–**k**, Results for a 2D molecular gas system, highly mobile molecules of fluorinated copper phthalocyanine ($F_{16}CuPc$) on a Si(111)/Tl-(1×1) surface. **f**–**i** panels correspond to Fig. 4a,e,b,f of Matvija et al[50], respectively. Copyright 2017 American Chemical Society. **f,g**, Experimental STM images of mobile $F_{16}CuPc$ molecules surrounding one (**f**) and two (**g**) fixed $F_{16}CuPc$ molecules. **h,i**, Molecular density maps from a kinetic Monte Carlo simulation, corresponding to the same systems shown in **f,g**. **j,k**, Molecular density maps calculated from our SLS model, corresponding to the same systems shown in **h,i**. Scale bars (**h**–**k**): 3 nm. In both the liquid helium and 2D molecular gas systems, SLS results are in agreement with existing results in the corresponding literature. Details of SLS calculation procedures for both systems are provided in Supplementary Note 7.